% ...

\documentclass[jcp,twocolumn,showpacs,preprintnumbers,amsmath,amssymb,floatfix,reprint]{revtex4-1}
\usepackage{hyperref}
\usepackage{epsfig}
\usepackage{color}
\begin{document}

\title{Squeezout phenomena and boundary layer formation of a model ionic liquid under confinement and charging}

\author{R. Capozza$^{1,2}$, A. Vanossi$^{2,1}$, A. Benassi$^3$, and E. Tosatti$^{1,2,4}$}
\affiliation{
$^1$ International School for Advanced Studies (SISSA), Via Bonomea 265, 34136 Trieste, Italy \\
$^2$ CNR-IOM Democritos National Simulation Center, Via Bonomea 265, 34136 Trieste, Italy \\
$^3$ Empa, Federal Laboratories for Materials Science and Technology, \"Uberlandstrasse 129, 8600 D\"ubendorf, Switzerland \\
$^4$ International Centre for Theoretical Physics (ICTP), Strada Costiera 11, 34014 Trieste, Italy
}

\begin{abstract}

Electrical charging of parallel plates confining a model ionic liquid down to nanoscale distances yields a variety of 
charge-induced changes in the structural features of the confined film. That includes even-odd switching of the structural 
layering and charging-induced solidification and melting, with important changes of local ordering between and 
within layers, and of squeezout behavior. By means of molecular dynamics simulations, we explore this variety of phenomena in the simplest 
charged Lennard-Jones  coarse-grained model including or excluding the effect a neutral tail giving an anisotropic shape to one of the 
model ions. Using these models and open conditions permitting the flow of ions in and out of the interplate gap, 
we simulate the liquid squeezout to obtain the distance dependent structure and forces between the plates during 
their adiabatic appraoch under load. Simulations at fixed applied force illustrate an effective electrical pumping 
of the ionic liquid, from a thick nearly solid film that withstands the interplate pressure for high plate charge to 
complete squeezout following melting near zero charge.  Effective enthalpy curves obtained by integration of 
interplate forces versus distance show the local minima that correspond to layering, and predict the switching 
between one minimum and another under squeezing and charging.        
 
\end{abstract}
\date{\today}
\maketitle


\section{Introduction}
Ionic liquids (ILs), ionic salts based on organic molecules whose large size, amphiphilicity and anion-cation 
asymmetry generally yield melting below room temperature, encompass an enormous chemical variety \cite{zhang},
interesting physical properties \cite{greaves06}, and numerous applications \cite{plechkova08}. 
The  structural and electrical properties of ILs near solid interfaces have been investigated experimentally \cite{mezger08,hayes11} 
and theoretically \cite{kornyshevtheory,kornyshev07}. 
Data show that under force-induced boundary confinement, the liquid
undergoes strong layering ~\cite{atkin07, bennewitz, perkin, li14}. Squeezout takes place for increasing force
in bilayer steps, so that ions of opposite sign exit together, keeping overall neutrality.
The electrical charge of the lubricant molecules and their layering under confinement naturally suggest the possibility to
influence the layering, and the squeezout behavior, by means of plate charging or by otherwise applied fields. 
The charging of a gold surface in an IL filled narrow AFM gap indeed showed important frictional changes as a function 
of the applied voltage \cite{bennewitz}. 
Different effects of charging have been reported in other experiments with different ILs ~\cite{li14}.
That confirms on one hand that charging is indeed an important parameter influencing structure, squeezout and friction,
and on the other hand that the details of these charging effects generally depend upon the specific 
nature of the IL and of plates.  
It seems at this point important from a more general viewpoint to conduct a  wider theoretical and
simulation inquiry, exploring possible charging induced phenomena based on less specific, more
generic, reasonably simple models.
These models of course do not replace the much more specific systems currently being simulated to describe specific 
ILs and their confineb behavior~\cite{padua13,federici14}.   
Simplicity of the model on the other hand is essential to allow a wide-angle view of some among 
the possible squeezout effects and scenarios provided by charging, either static or time-dependent, 
of the confining surfaces. 

Within this broad exploratory program, and without aiming at modeling a specific case,
we investigate first of all the scenario of ILs confined between two neutral
or nearly neutral, then charged, solid plates. This study encompasses a  variety of charging-related static phenomena,
preparatory to the dynamical study of frictional behaviour which will be considered in subsequent work.\\ 
As in the early approach of Fedorov and Kornyshev we begin with a simple charged Lennard-Jones system, 
but then we extend it by the minimal  addition of a neutral ``tail'' making one of the ions considerably 
asymmetric, as is common in many real IL ions.
Using an open geometry, where ions can escape or re-enter the interplate gap during slow, adiabatic
changes of the interplate gap width $D$, we observe in the simulation the formation of partly ordered layers 
under squeezing. In analogy to experiments~\cite{hayes11, perkin} squeezout  is found to take place by successive transitions
through plateaus at well defined interplate distances, through force-induced gap changes corresponding 
to one IL bilayer, clearly the neutral entity here. This transverse layering structure is accompanied by some degree of lateral, 
planar ordering inside each IL molecular layer. Planar order, even if probably overemphasized by our simple
models relative to much more complex real ILs (where in-plane spatial order is so far unaccessible experimentally)
is nonetheless a relevant feature that helps understanding both structure and dynamical behavior. We characterize 
planar ordering of our confined IL by means of a k-space resolved structure factor, analogous to a two-dimensional, 
z-resolved diffraction scattering amplitude. 
Some level of frustrated plate- and confinement-induced 3D crystallization is also found, whose detailsare, as we show, 
model-dependent. We extract by means of force-distance integration a very instructive effective interplate interaction 
free energy, where the different plateaus appear as local minima, so that the squeezout transitions between 
them is described and in fact predicted by a Maxwell construction.  
The  effect of charging is subsequently shown to cause structural changes
which can in turn be related to changes of the interplate equilibrium free energy
characteristics. 

\section{Model and simulation geometry}
\begin{figure}
\includegraphics[width=1.\linewidth]{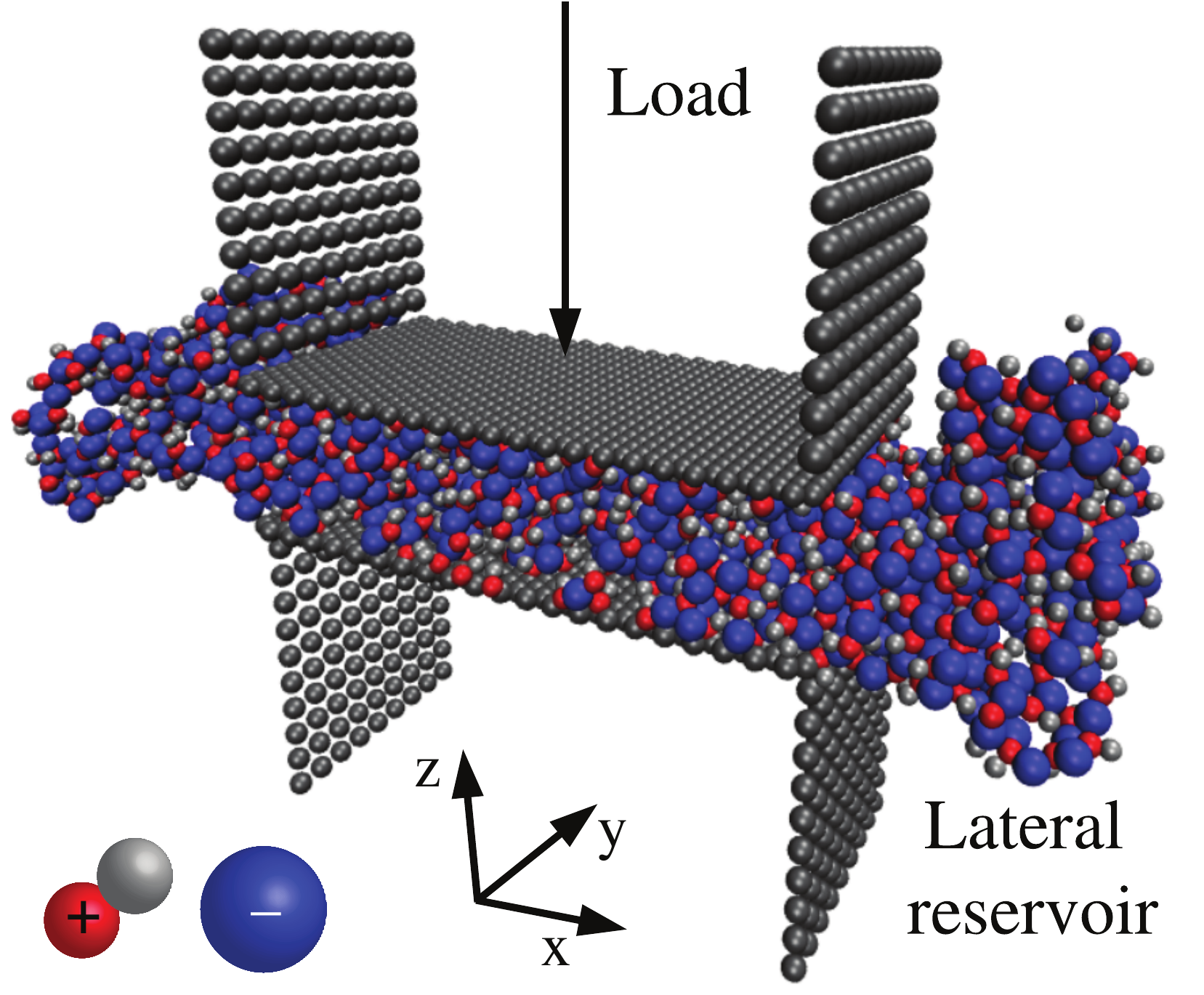}
\caption{(Color online) Simulation geometry with open boundaries along $x$ and $z$ directions
and periodic boundaries along $y$, schematically similar to SFA. The x size of plates is $20nm$.
By application of a load to the top plate, the liquid can flow to lateral reservoirs. When the load is reduced
or reversed, the liquid flows back inside the gap. In the lower left corner the anion and cation shaped are sketched. The cation may not or may have an attached neutral tail, as indicated. }
\label{sketch}
\end{figure}

ILs are molten salts usually made up of large-size anions and cations \cite{zhang}, generally organic,
and with  asymmetric irregular shapes often including long alkyl chains. The irregularity is
important as it gives the molecule a larger gyration ratio in the liquid state, while effectively
preventing low temperature order and crystallization by replacing it with a glassy state. 
Dropping most of these complications we restrict to a much simpler model, crudely including just  
the rough features of a generic IL. We will therefore build on the basic model already exploited 
in previous studies~\cite{kornyshev}, namely a charged  Lennard-Jones (LJ) system where aniona and cations have different radii. 
We compare this model with its extension, obtained by attaching a neutral "tail" to the cation. In the extended model  
the anion consists of a negatively charged, large-sized spherical LJ classical particle, the cation of a  dimer 
made of a positively charged, small-sized LJ head, rigidly bound to an equal size, neutral particle, the tail 
(see Fig. \ref{sketch}). The tail simultaneously achieves several goals. First, it enhances the liquid gyration
diameter, the solid-liquid density jump, and the general tendency to form a glass rather than a crystal at low temperatures.
Second, it improves the plate wetting behavior, which without tails is unrealistically strong, while the wetting 
by real IL of surfaces such as mica is only partial \cite{wang}. Partial wetting is plausibly 
associated, at least in some cases, with a first monolayer of cations whose outward pointing tails may provide 
a ``phobic'' gap against further layer-by-layer film growth. That is realized  in 
our ``tailed'' model (TM) but not in the simpler charged LJ "salt-like" (SM) model. Finally
the tail also introduces rotational entropy effects that must be present in real systems, 
but are missing in simple SMs \cite{kornyshev,fumi,alejandre}.\\

Building on the work of Fedorov et al. \cite{kornyshev}, we thus assume the anion (A), 
the positive head of the cation (C) and plate "atoms" (P) to interact electrostatically 
and via a LJ potential 
\begin{equation}
V(r_{ij})=4\epsilon_{ij}\left[\left(\frac{\sigma_{ij}}{r_{ij}}\right)^{12}-
\left(\frac{\sigma_{ij}}{r_{ij}}\right)^6 \right]+\frac{e_i e_j}{4\pi\epsilon_0\epsilon_r r_{ij}}
\label{ljcharge}
\end{equation}
with $i,j=A,C,P$. 
The tails (T) interact with all other species through a repulsive potential
\begin{equation}
V^T(r_{lT})=4\epsilon_{lT}\left[\left(\frac{\sigma_{lT}}{r_{lT}}\right)^{12}\right]
\label{ljrep}
\end{equation}
with $l=A,C,P,T$. The full list of parameters is reported in Table \ref{table}.
We adopt the LJ potential strengths $\epsilon_{AP}=\epsilon_{CP}=K_B T_{room}/2=13.3 meV$ 
as our energy unit and take ion-plate radii slightly smaller than those between ions in order 
to provide a reasonable wetting habit of the plates.
The SM consists of the same model liquid simply without the tail, as used in ~\cite{kornyshev}.
To account for the dielectric screening of real systems, the two ionic species of charge +$e$ and -$e$ ($e$ 
the electron charge) are immersed in a uniform  average dielectric with $\epsilon_r =2.0 $. 

\begin{table}[h!]
\centering
  \begin{tabular}{ l | c | r }
    \hline
     & $\sigma(nm)$ & $\epsilon(meV)$ \\ \hline
    AA & 1 & 1.3 \\ 
    CC & 0.5 & 1.3 \\ 
    AC & 0.75 & 1.3 \\
    AP & 0.7 & 13 \\
    CP & 0.35 & 13 \\
    TT & 0.5 & 1.3 \\
    TA & 0.75 & 1.3 \\
    TC & 0.5 & 1.3 \\
    TP & 0.35 & 13 \\ 
\end{tabular}
\caption{List of LJ parameters used in the simulations.}
\label{table}
\end{table}

The liquid is confined between two negatively charged plates with a modest surface charge density $q=-4\mu C/cm^2$, unless 
differently specified, chosen to break the charge symmetry. The plates (somewhat similar to mica surfaces used in SFA experiments) 
are made of LJ sites arranged in a rigid close-packed triangular lattice with spacing 0.52 nm.\\All 
molecular dynamics (MD) simulations were performed using the LAMMPS code~\cite{lammps}. 
Canonical ensemble configurations were sampled by means of a Langevin thermostat 
directly applied to the lubricant molecules. The plates were treated as rigid bodies, 
the lower one fixed and the upper one either subjected to a z-directed force $F_n$ (the load) as 
shown in Fig.~\ref{sketch}, or else driven along $z$, inwards or outwards, at some very small constant velocity. 
While the total number of simulated ions is constant, the finite plate width along $x$ ending with free vacuum on both sides
in the simulation cell geometry (Fig.~\ref{sketch}), permits, unlike other IL simulations, otherwise
very realistic,~\cite{federici14,padua13} particle squeezout with formation of two lateral ionic liquid
drops. The drops serve as liquid reservoirs so that the number of ions effectively confined inside 
the gap can  dynamically change depending on the loading conditions, realizing an effectively grand-canonical situation.\\

\section{Bulk melting and solidification}

\begin{figure}
\centering
\includegraphics[width=1.\linewidth]{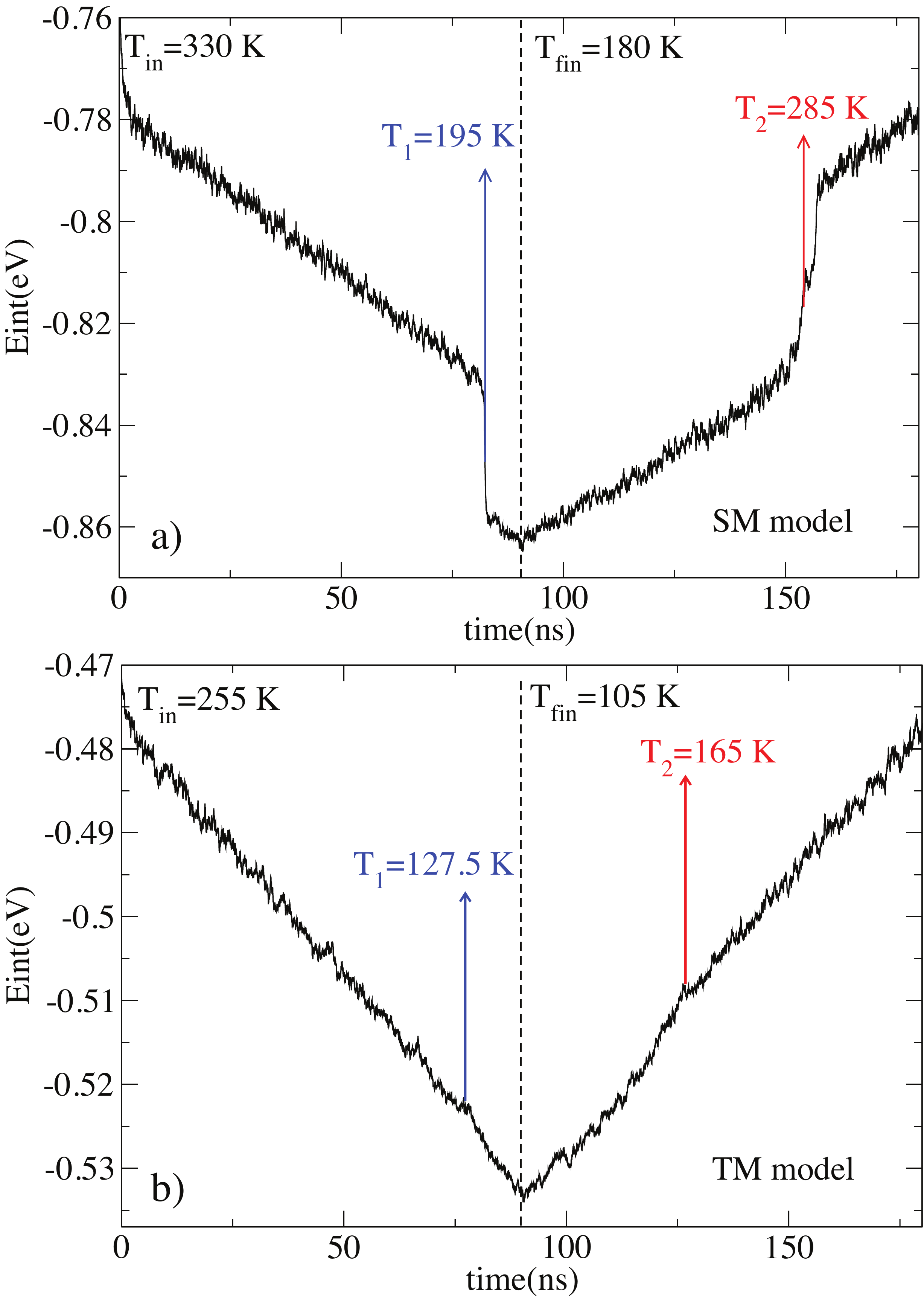}
\caption{(Color online) Bulk internal energy $E_{int}$ of the SM (upper panel) and TM (lower panel) liquids as a function of simulation time,
during a linear back and forth temperature ramping of rate $1.7K/ns$.}
\label{int_en}
\end{figure}

\begin{figure}
\includegraphics[width=1.\linewidth]{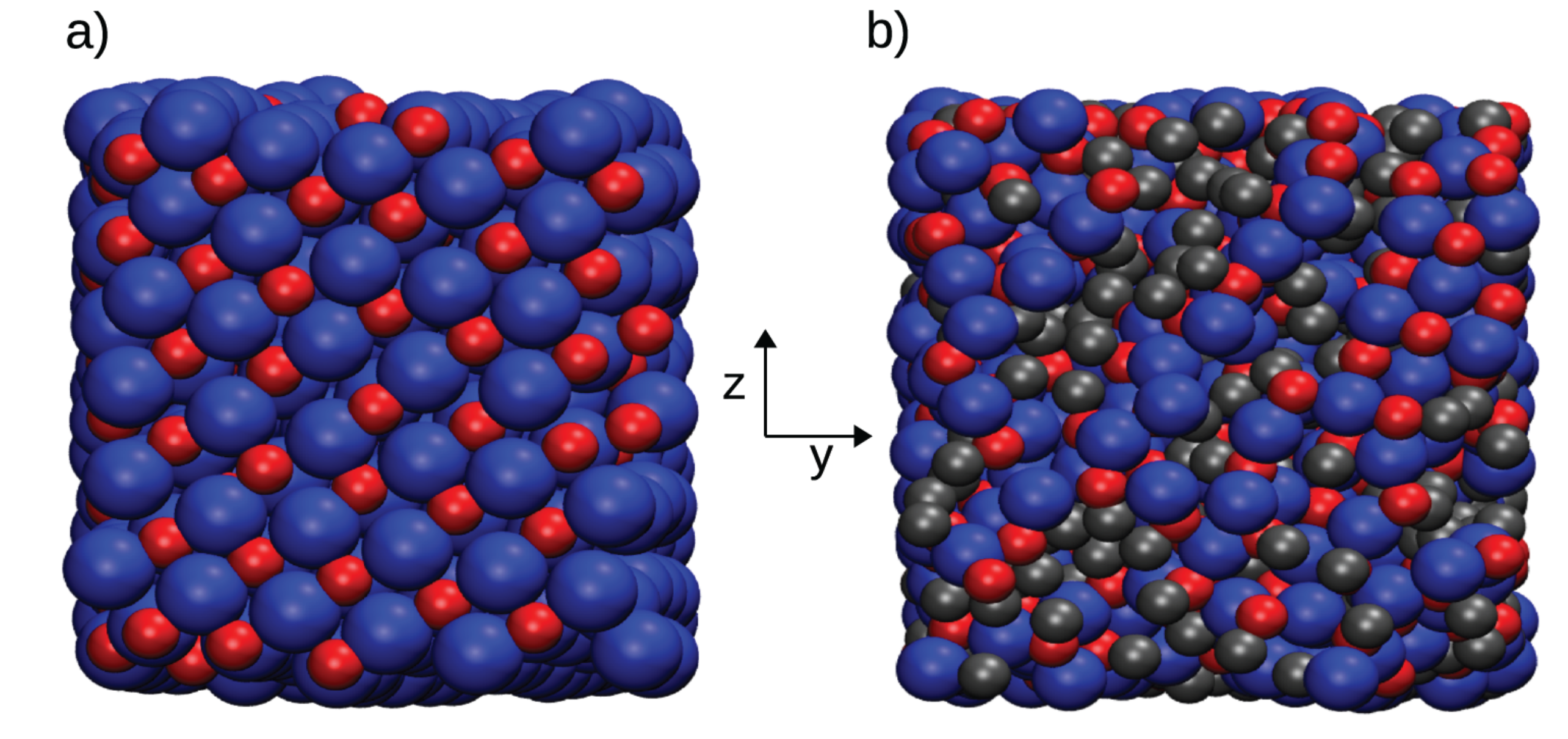}
\caption{(Color online) Snapshot of (a) SM  and  (b) TM 
bulk solids below their respective melting temperatures. 
The presence of tails in TM prevents crystallization in the rocksalt structure  as in SM.}
\label{solid-tail}
\end{figure}

A basic property of an IL is its melting habit. We 
investigate and compare the bulk melting temperature of the TM and SM models by performing
a slow-varying temperature loop MD simulation of both models with $N=2000$ ions and full 3D periodic 
boundary conditions. Starting from a  high temperature $T_{in}$, where both SM and TM  
are in the liquid state, we  
first decrease the temperature to 
a much lower $T_{fin}$ with a rate of $1.7K/ns$, producing solidification;  then increase it again back 
to $T_{in}$, to produce melting .
Fig.\ref{int_en}a shows the internal energy, $E_{int}$ of the SM inthe temperature loop.
On cooling, $E_{int}$ linearly decreases
until a sharp drop at $T_1=195K$ corresponding to a liquid-solid transition. 
On heating back the solid a sharp rise of $E_{int}$ is found at $T_2=285K$, corresponding
to the reverse solid-liquid transformation. The large magnitude of jumps reflects strong crystalline order
in the SM solid. The equilibrium melting temperature is to be foumd between those of internal energy drop 
and rise, whose difference reflects the hysteresis normally encountered in the absence of nucleation centers, such as surfaces.
The TM behaves similarly, but displays smoother transitions, reflecting a much poorer ordering in the solid, 
and lower transition temperatures 
$T_1=127.5K$ and $T_2=165K$, as shown in Fig.~\ref{int_en}b.\\ 
In both SM and TM these transitions are 
accompanied by correspondingly strong changes of diffusion coefficient (not shown). If we assume arbitrarily that
the melting temperature is the average between $T_1$ and $T_2$, we obtain $T_{TM}^m\simeq 150 \pm 20 K $ 
for TM and  $T_{SM}^m\simeq 240 \pm 40 K$ for SM respectively. The decreased density 
of TM relative to SM reduces the cohesive energy and thus the melting temperature. The tails in TM also
prevent crystallization in the rocksalt structure typical of SM as shown in Fig.\ref{solid-tail}.
Unless otherwise specified, all simulations reported in the following were done at 
temperature $T_{TM}=225 K$ and $T_{SM}=300 K$ for the TM and SM respectively, safely above
but not too far above the bulk melting temperatures, as in real experimental cases. In our simulations
therefore both ILs are in a fully liquid state as long as the gap width between the two plates is sufficiently  large. 
Both ILs may solidify, as we will show, when the confining gap between plates is reduced down to the 
molecular thicknesses typical of boundary lubrication conditions.  


\section{Wetting properties}

\begin{figure}
\includegraphics[width=1.\linewidth]{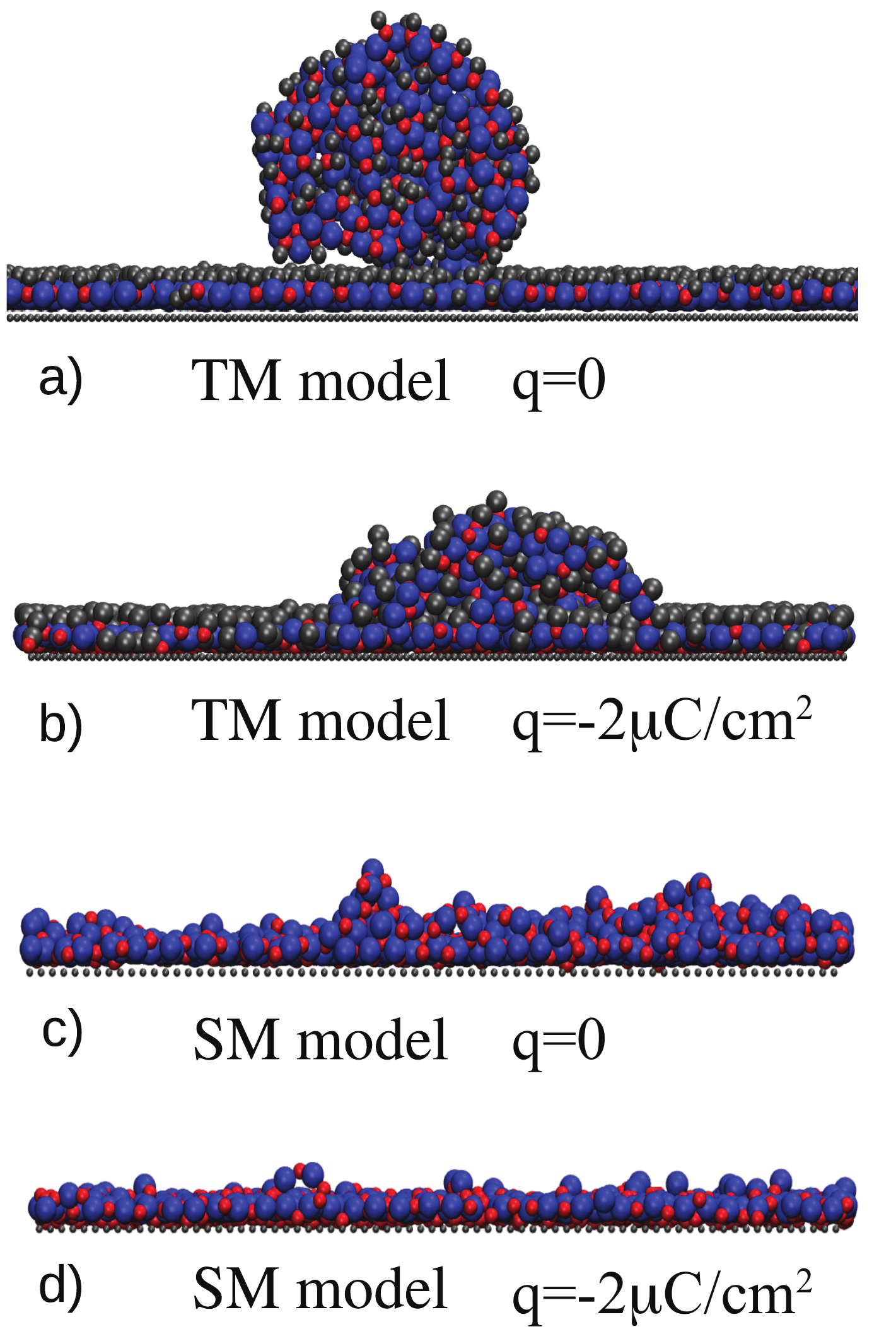}
\caption{(Color online) Simulated wetting behavior of a neutral and a moderately negative plate 
by  an IL droplet in (a, b) TM and (c,d)  SM models. Both SM and TM coat the plate with a cation monolayer. 
In the SM liquid that is followed by deposition of subsequent full layers indicating complete wetting. In the TM liquid 
the cation monolayer provides a neutral, ``phobic'' coating preceding the formation of a droplet, indicating partial wetting. 
Plate charging causes in this case a drop of wetting angle from above 90$^{\circ}$ to $\sim 30 ^{\circ}$ (electrowetting). The
simulation temperatures are $T_{TM}=165K$ and $T_{SM}=300K$ for the TM and SM respectively.}
\label{drop}
\end{figure}

The wetting of plate surfaces such as mica is known to be partial by at least some 
ILs \cite{beattie2013,wang}. Squeezout phenomena and layering necessarily involve 
intimate features of the liquid-solid plate interface, 
not disconnected with those determining the plate wettability. Thus it is important to monitor the plate wetting habit of our  IL models. 
To do that, we simulate for both models the behaviour of a liquid droplet 
consisting of $N=2000$ ions
deposited on a neutral and then on a lightly charged plate. The adhesion parameters of 
the liquids should be chosen in a reasonable way and in order to reproduce qualitatively  
the wetting properties. We find (Fig.\ref{drop})  that the TM wetting of the neutral plate is poor, forming a wetting angle $\theta>90^{\circ}$. 
On the contrary, the SM completely wets the substrate. Only in the TM the cations provide 
outwards pointing tails, which give rise to a "phobic" gap, responsible for the incomplete wetting, see Fig.\ref{drop}a.
Upon charging of the plates, one expects wetting to be generally enhanced \cite{welters98,paneru10}
for either sign of charge, over the neutral case. 
For moderately negative charging  we find that the TM liquid indeed 
wets better than in the neutral case, yielding a smaller wetting angle $\theta\simeq30^0$.  
The charging evidently reduces the effect of tails, and their effective screening of electrostatic forces.
For a moderatly positive charge on the other hand, the TM wetting angle $\theta$ is slightly
larger than negatively charged substrate. This asymmetry is due to the different size and shape of ions.\\    
These TM results -- monolayer coating, partial wetting by the drop -- are in qualitative agreement 
with experimental data \cite{beattie2013,wang} for 1-butyl-3-methyl-imidazolium bis-(triﬂuoromethylsulfonyl)imide, 
[BMIM] [TFSI], on mica, 
which show a precursor layer forming ahead of a macroscopic drop with $\theta \simeq 30^{\circ}$.
This partial wetting scenario is absent, and replaced by complete wetting in the SM.
Thus, in such simplified modeling, the introduction of neutral repulsive tails, 
mimicking, e.g., alkyl chains in real IL systems, turns to be important for the correct wetting properties of the IL.
It is highly reasonable to presume that this important element of realism will be important in future
studies of sliding friction. \\
\section{Squeezout simulations and layering}

Liquids close to hard surfaces display, both experimentally 
and theoretically,  spatial structuring phenomena with respect to their uniform bulk density. When a liquid film is confined 
in a planar gap between hard plates a few molecular diameters wide, 
it develops quite generally a  layered density profile \cite{israelachvili11}, with increasing solid-like properties that eventually 
enable it to resist squeezout and to support static friction. The squeezout of these layers one at a time 
under increasing load is well studied theoretically~\cite{persson94, gao97b, tartaglino02,tartaglino06} 
and experimentally~\cite{mugele00,zilberman01}.\\
Similar to common liquids, ILs  generally give rise to partially ordered molecular-size layers and 
enhanced viscosity at interfaces with solid surfaces such as mica, silica, and gold \cite{ueno10,smith13,atkin07,atkin09}. 
For reasons of charge neutrality and compatibly with ion asymmetry, ILs organize themselves in alternating positive and negative layers 
not too far from the plates. 
The squeezout between approaching plates occurs in this case by neutral entities, that is by pairs of layers, 
as seen in all experiments, as also described by a recent theoretical model~\cite{hoth14}.
Depending on the specific IL, its melting temperature and its relationship to 
the confining surfaces, experimental 
squeezouts by sharp AFM tips may require relatively large applied forces from some nN 
to several tens of nN \cite{hayes11} which, for an area in the order of 1-10 nm$^2$ 
gives an indication of the strength of interactions involved.\\ 

\begin{figure}
\includegraphics[width=1.\linewidth]{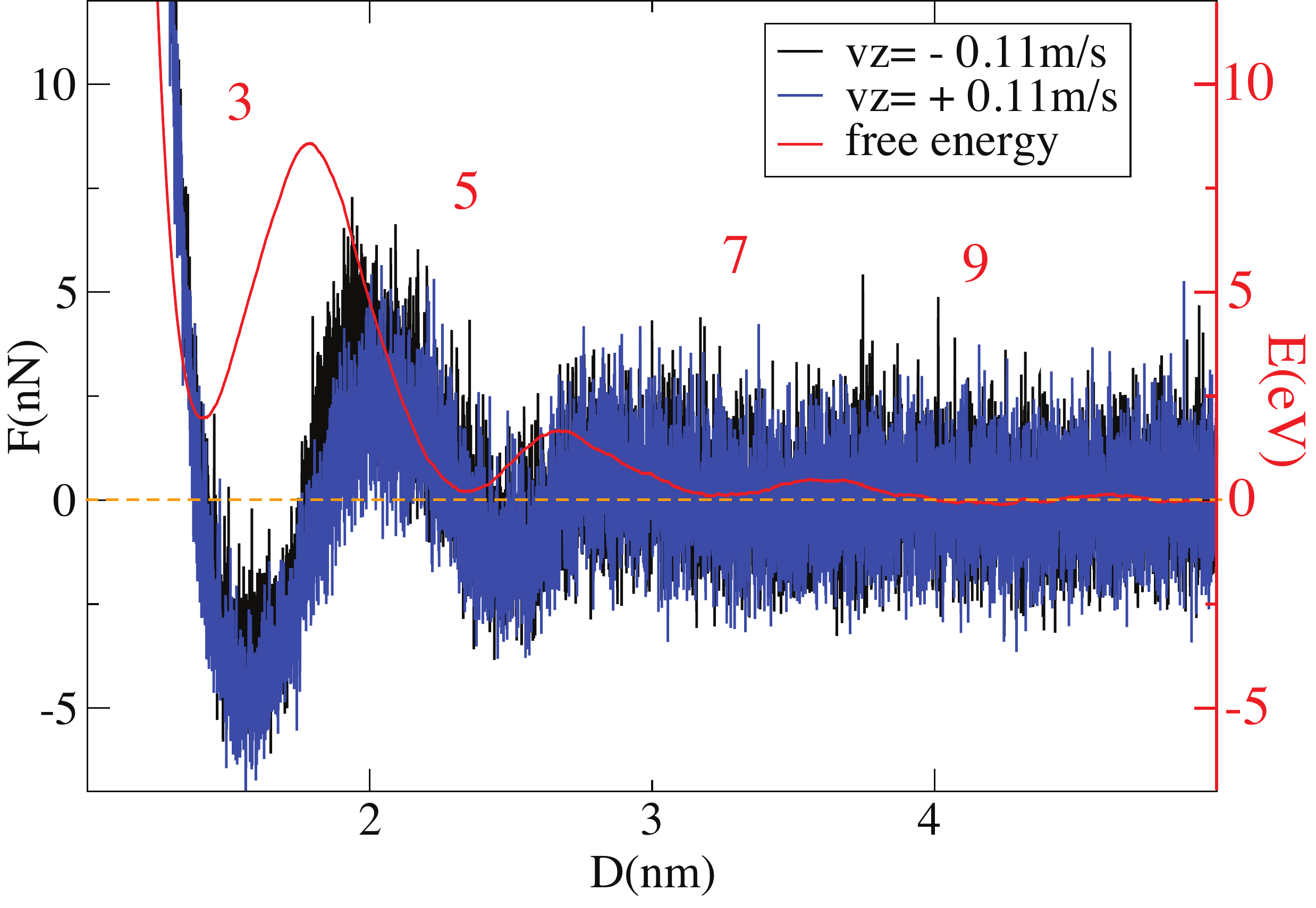}

\caption{(Color online) Force $F$ as a function of distance $D$ between plates.
Black and blue curves correspond to downward and upward motion of the top plate
respectively. The force difference is due to the small but finite plate speed $vz=0.11m/s$,
and their average represents our best approximation to the adiabatic force $F(D)$. 
The red curve is the free energy $W(D)$ obtained by integration of  the adiabatic 
force. Numbers indicated in figure denote the number of IL layers confined between 
the plates. The charge on both plates is $q=-4 \mu C/cm^2$, a modest value compared
with that of e.g., mica~\cite{atkin07}.}
\label{force_dist}
\end{figure}

The available simulations of confined ILs are carried out under unrealistic sealed conditions, 
via periodic boundary constraints \cite{padua13,federici14}, which do not 
permit squeezout. Experimental SFA and AFM setups constitute eminently open geometries, 
allowing dynamical liquid squeezout and suck-in.
For our simulations we therefore adopt an open geometry, albeit a very simple and schematic one. 
As shown in Fig.\ref{sketch} it consists of two rigid plates that are infinite along $y$, but finite along $x$,
with the IL filling the gap between them.  When the plates are moved closer to one another, 
the liquid is squeezed out of the gap,  flowing sideways to form two side drops that constitute IL reservoirs. 
A simulation geometry similar to ours was used by Landman's group ~\cite{gao97b}
whereas a different open geometry allowing for squeezout was adopted  by Tartaglino et al.
~\cite{tartaglino06}. Open simulations were also recently used to investigate squeezout and phase transition of an argon 
film between two solid surfaces \cite{leng13}. 
In our initial protocol the plates were very slowly approached, and the resulting average force 
on the top plate, $F$, was monitored as a function of the diminishing inter-plate distance $D$.  
Beginning for example with a TM bulk liquid  confined between plates and closing gradually the 
gap width $D$, $F$ remains initially zero while the liquid is readily expelled, until a critical distance is reached 
between plates, $D_c\simeq 4nm$,  amounting to a few bilayers, 
where the force-distance curve departs significantly from zero and begins to grow. The growing force $F(D)$ 
between the plates shows layering oscillations as in Fig.\ref{force_dist}. 
Each force dip corresponds to the complete squeezeout of a pair
of layers, one positive and one negative, as in experiment~\cite{hayes11,perkin}. 
If the plate motion is inverted, now increasing the gap width with time, the squeezout 
is replaced by "suck-in" of the IL. The black and the blue force 
curves of  squeezout and suck-in 
are not identical at a given distance, and form a narrow hysteresis loop. 
The area enclosed in the force-displacement loop measures the dissipation 
work implied by our plate to-and-from motion, exerted at small but not 
infinitesimal speed. Were the plate speed to tend to zero (quasi-static limit), dissipation 
would vanish, and the two force-distance curves would collapse onto a single 
adiabatic force $F(D)$ at all finite temperatures. 
In these conditions the adiabatic work $ W(D)  = \int_D^\infty F(z) dz$ 
provides an accurate measure of the interaction free energy between the plates. Layering oscillations of 
this distance-dependent free energy was earlier demonstrated in simulations of simple liquids~\cite{gao97b}.

\begin{figure}
\includegraphics[width=1.\linewidth]{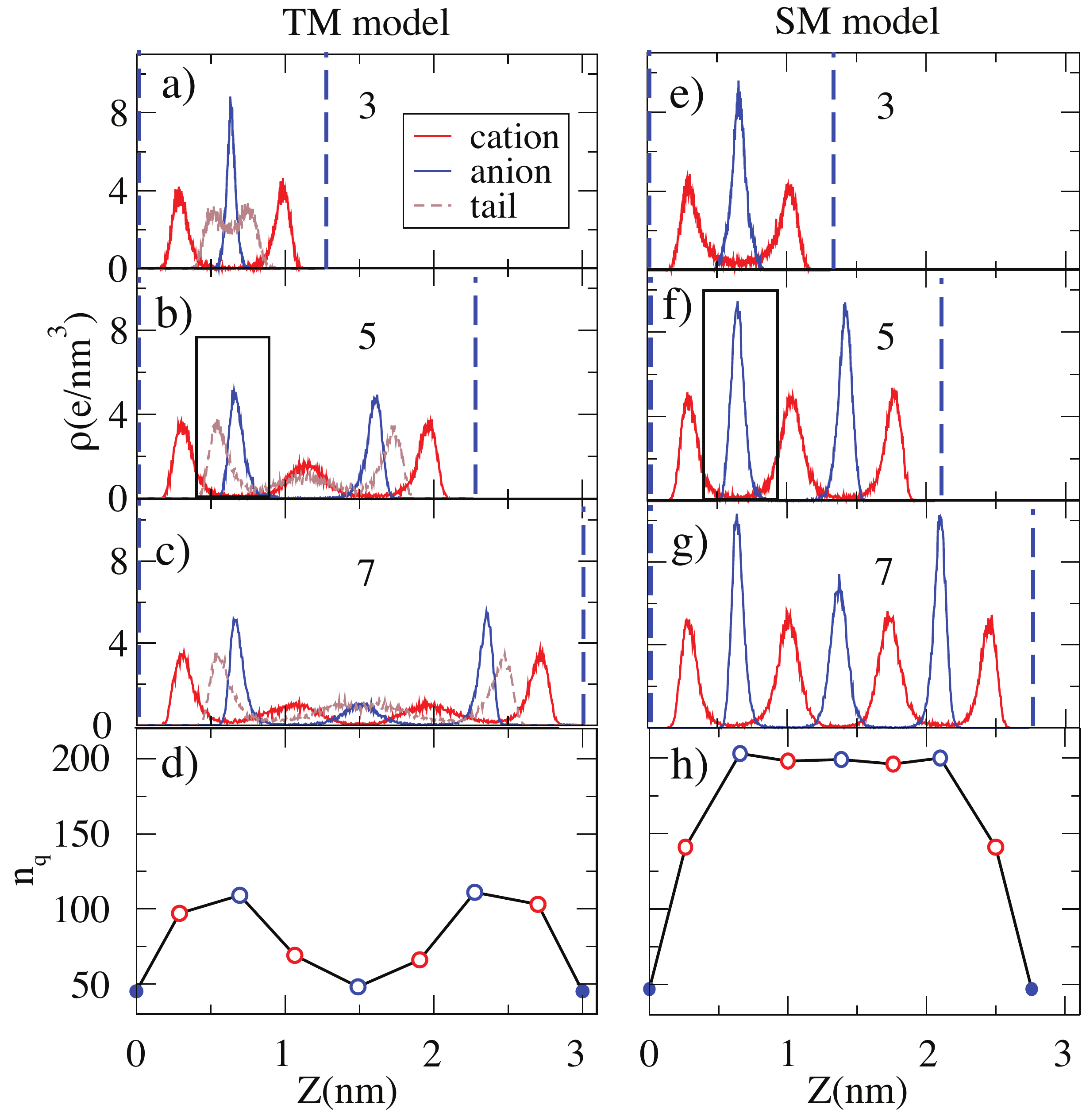}
\caption{(Color online) Density profile along $z$ axis for the TM (left) and SM 
liquid (right) and increasing numbers of confined layers. 
Vertical blue dashed lines indicate the position of the plates at separation $D$, light brown peaks in panels 
a), b), c) show the density of tails. Black squares in panels f) and b)
mark the single layers that are analyzed in Fig.\ref{sqsalt}b and \ref{sqtail}b.
Lowest panels: number of charges per layer, $n_q=Q_l/e$, with $Q_l$ total charge per 
layer and $e$ electron charge.
The full circles at the edges of panels indicate the charge assumed on the plates.}
\label{vprof}
\end{figure}

Changing from fixed distance to fixed force, where a normal load, $F_n$, is applied to the top plate,
and the interplate distance $D$ is observed to vary as a consequence, these free energy curves establish a relation between 
$F_n$ and the consequent number of confined layers. For example, 
when $F_n>3.5nN$ only two metastable states with 3 or 5 layers are allowed, while at $F_n\simeq 0$ all states are allowed.
That concept was recently underlined by Hoth et al.~\cite{hoth14}.\\ 
The number of confined layers between identical plates is odd, as is demanded by symmetry. 
Confinement of ILs between mica plates offers a direct example of odd layering \cite{perkin} 
due to the symmetrically slightly negative charge of both plates, each of which attracts a similar coating
of cations. The "transverse" density profile measured along the vertical, plate-to-plate z axis shows a clear 
oscillatory order with cations and anions populating different, well separated layers. In the TM in addition, the 
light brown line represents the density of tails (Fig.\ref{vprof}a,b,c). The positions of cation and anion peaks closest 
to the plates (vertical blue dashed lines), are geometrically related to their radii. As the number of layers increase, 
all oscillations broaden and weaken, more so in the TM due to the disordering effect of tails.
The SM liquid presents a qualitatively similar picture, where the absence of tails and related disorder 
allows much more pronounced layering order, occurring at a larger distance between the plates. 
The force needed to squeeze out a pair of SM layers is much larger, and the peaks of density 
profile along vertical axis more pronounced, as shown in Fig.\ref{vprof}e,f,g.\\

The bottom panels in Fig.\ref{vprof}d,h display, in the case of seven layers, 
the number of charges per layer, $n_q=Q_l/e$, 
with $Q_l$ total charge per layer and $e$ electron charge. 
The total  amount of charge, including the plates and the confined liquid, 
preserves electroneutrality.
The TM shows a clear overscreening, with the charge of layers in
contact with plates substantially larger than far away. 
Due to confinement, the SM liquid shows a much readier tendency to solidify.  
The charge plateau in Fig.\ref{vprof}h indicates precisely a 
crystalline region where a well defined number of ions per layer 
is required for a rocksalt structure with (111) planes parallel 
to the plates. 
%
\section{Layer-resolved planar structure, 2D structure factors; 3D structures, crystallization phenomena}
\begin{figure}
\includegraphics[width=1.\linewidth]{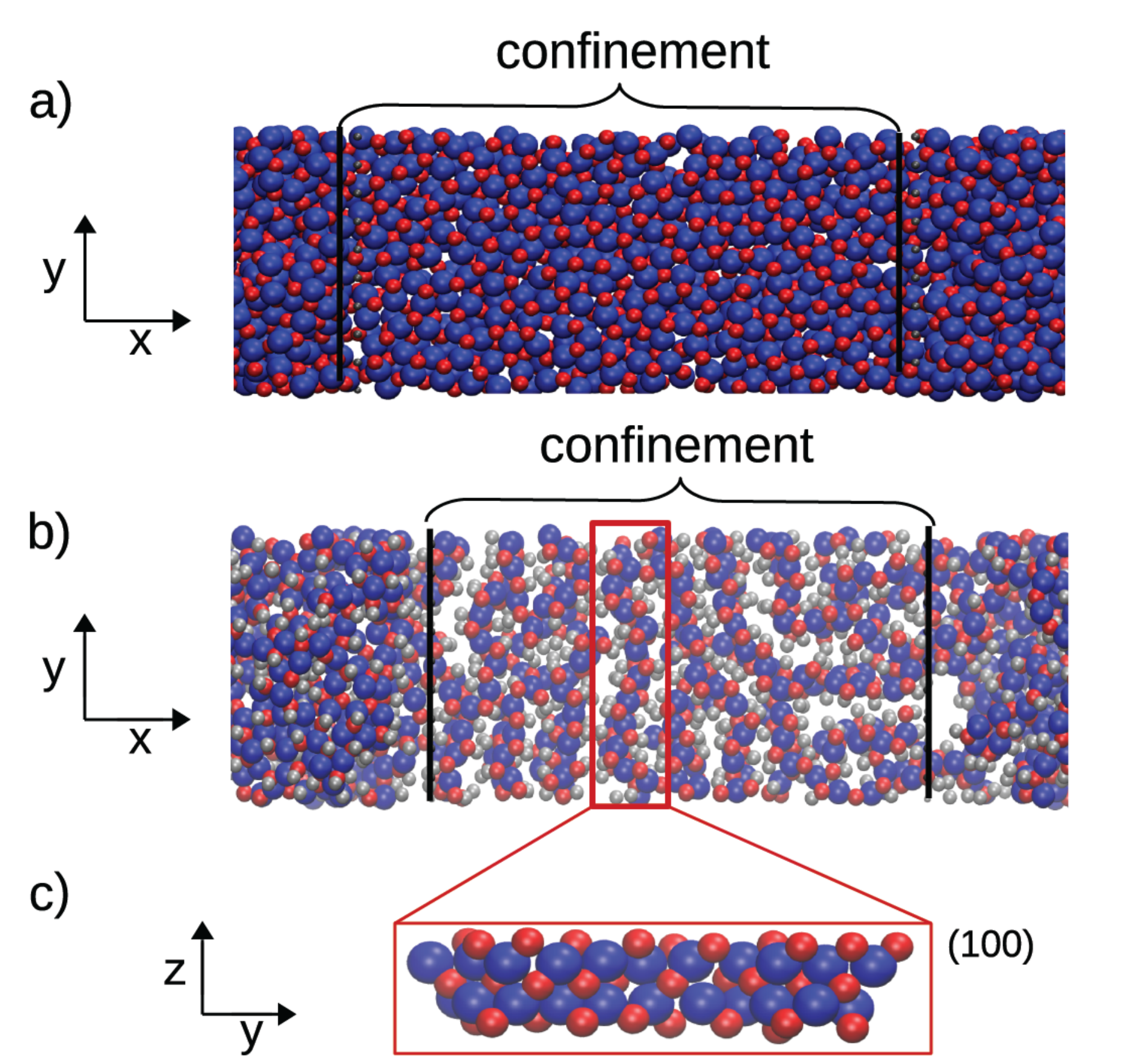}
\caption{(Color online) a) Snapshot of all SM particles for 5 layers, visualized looking down the z-axis.
 b) Same snapshot for the TM case. The meandering vertical walls are specific to TM and are absent in the SM.
 c) Side view of the single 2D meandering wall enclosed in the red rectangle of panel b). Tails are removed for clarity.}
\label{sqsalt}
\end{figure}

The simulation temperatures, substantially but not excessively above the bulk melting temperature,  ($T_{TM}=225K$ 
and $T_{SM}=300K$ for TM and SM respectively) preserve a reasonably large amount of 
short range order in the IL. That short range order naturally emerges near hard plates, and yields both transverse {\it and} planar 
static order in the layered structure of the liquid confined between the plates. That order is
associated with some partial solidification, and indeed  when the gap width $D$
decreases below a distance corresponding to 10 molecular diameters, both SM and TM layered 
ILs increasingly resist squeezout. Accompanying the layering, there must be some amount of planar ordering.
Contrary to layering, well known and clearly reflected by squeezout experiments, the planar order
is so far experimentally undetected.
Fig.\ref{sqsalt}a and \ref{sqsalt}b show top views of SM and TM liquids for $D$ corresponding
to five confined layers, revealing the nature and extent of planar order. 
The SM liquid has ordinary molten salt short-range order, eventually tending to (111) rocksalt planes parallel
to the plates. The effect of tails on planar order is quite strong.  Unlike the SM, the TM liquid forms 
wall-like structures, which transversely straddle the gap between
the plates. Ions of opposite charge arrange in charge-ordered planes {\it parallel} to z,
with wide gaps between them occupied by the tails, randomly protruding on either sides. While keeping
reasonably parallel at short range, these walls are rarely straight, and generally tend to meander as 
the figure shows. Similar wall-like (sometimes described as worm-like) patterns of unspecified heigth 
were reported in AFM pictures of ILs on gold \cite{atkin09, borisenko06,lin03} and mica \cite{segura13}. 

\begin{figure}
\includegraphics[width=1.\linewidth]{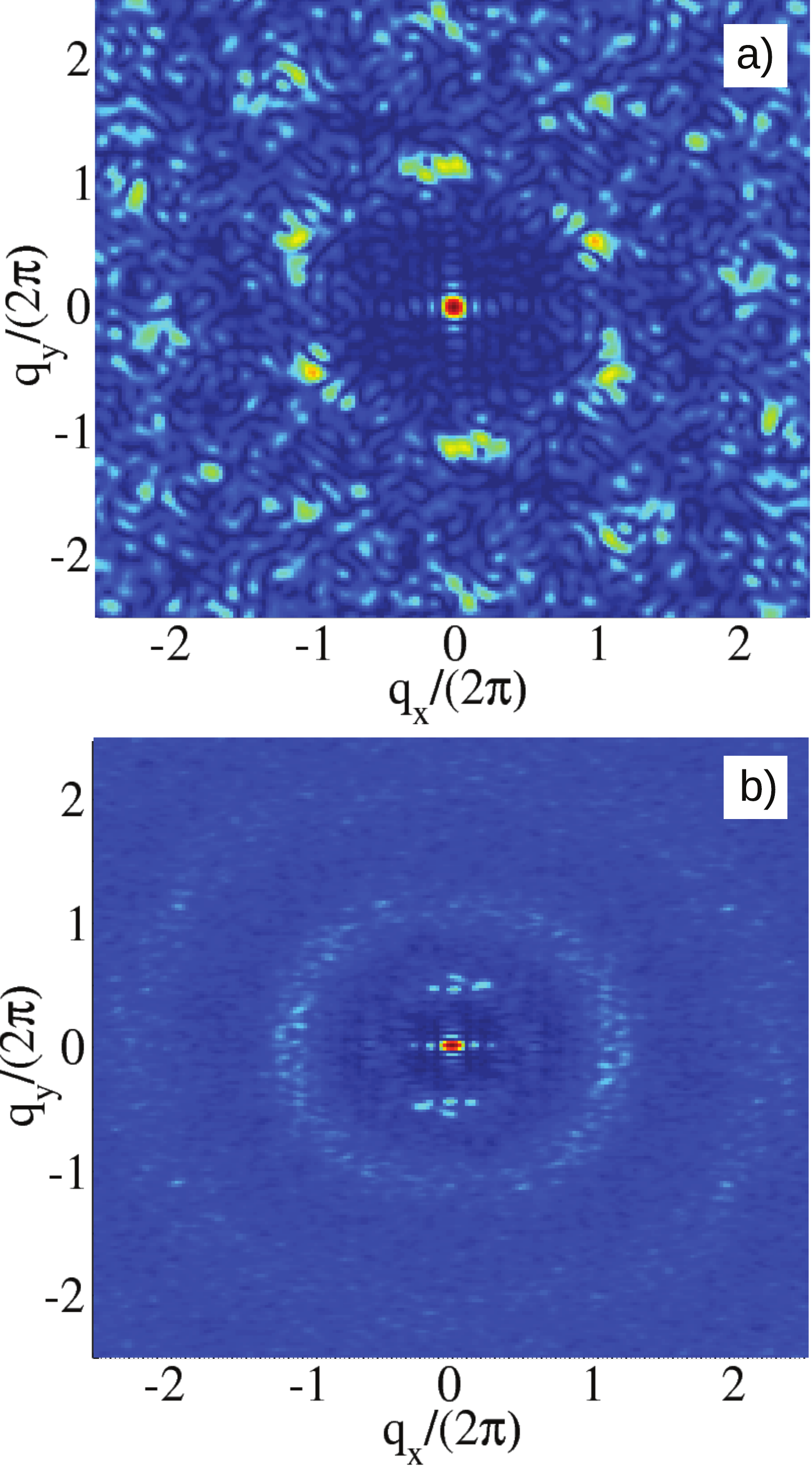}
\caption{(Color online) a) Planar structure factor $S_{z_0}(\mathbf{q})$ of the SM anion layer 
enclosed by the black rectangle  in Fig.\ref{vprof}e. b) Planar structure factor $S_{z_0}(\mathbf{q})$ of 
the TM anion layer enclosed by the black rectangle in Fig.\ref{vprof}b.}
\label{sqtail}
\end{figure}

For a more precise characterization of planar order, we calculated the 
z-resolved structure factor 
$S_{z_0}(\bf{q})$ per layer, defined as
\begin{equation}
S_{z_0}(\bf{q})=\frac{1}{N}\left\langle \sum_{j,k,z\in z_0}e^{-i\bf{q}(\bf{r}_j-\bf{r}_k)}\right\rangle
\label{sfactor}
\end{equation}
where $\bf{r}$ is the $(x,y)$ position of particles with $z$ coordinate
in a window close to $z_0$, $N$ is the number of particles in
that window, and $\langle...\rangle$ indicates time average. The z-resolved $S_{z_0}(\bf{q})$ of the window
enclosed in the black rectangle in Fig.\ref{vprof}f shows that anions approximately form a 2D 
triangular lattice, indicating that the SM liquid crystallizes in a rocksalt structure with a (111) 
plane in contact with the plate (see Fig.\ref{sqtail}a).
The confined TM  film behaves in this respect quite differently from SM. The layer structure factor $S_{z_0}(\bf{q})$ of the TM anion layer enclosed 
in the square of \ref{vprof}b, displays a liquid-like structure factor, caused by random 
orientation of wall-like structures (see Fig.\ref{sqtail}b).
Away from the plates the layering is less pronounced, and there the TM liquid 
shows a tendency to appear organized with a 2D triangular lattice symmetry.
Fig.\ref{sqsalt}c shows a side view of the single vertical wall 
indicated by the red square in Fig.\ref{sqsalt}b, where tails have been 
removed from the picture for clarity. The (100) plane of rocksalt is easily recognized. 
Because of the random orientation of tails, the liquid cannot solidify in a complete three-dimensional 
crystal, but it manages to arrange in vertical neutral rocksalt walls with tails segregated 
in the spaces between adjacent walls and oriented parallel to the plates. \\

These two examples suggest some more general understanding, that may be valid beyond the two specific
cases. First, plate-induced layering and charge neutrality are universal, as was to be expected. Second,
the detailed nature of planar and of three-dimensional ordering, ranging from nearly crystalline to
glassy, is variable and much more dependent upon the specific nature of the IL consituents.    

\section{Symmetric plate charging}

\begin{figure}
\includegraphics[width=1.\linewidth]{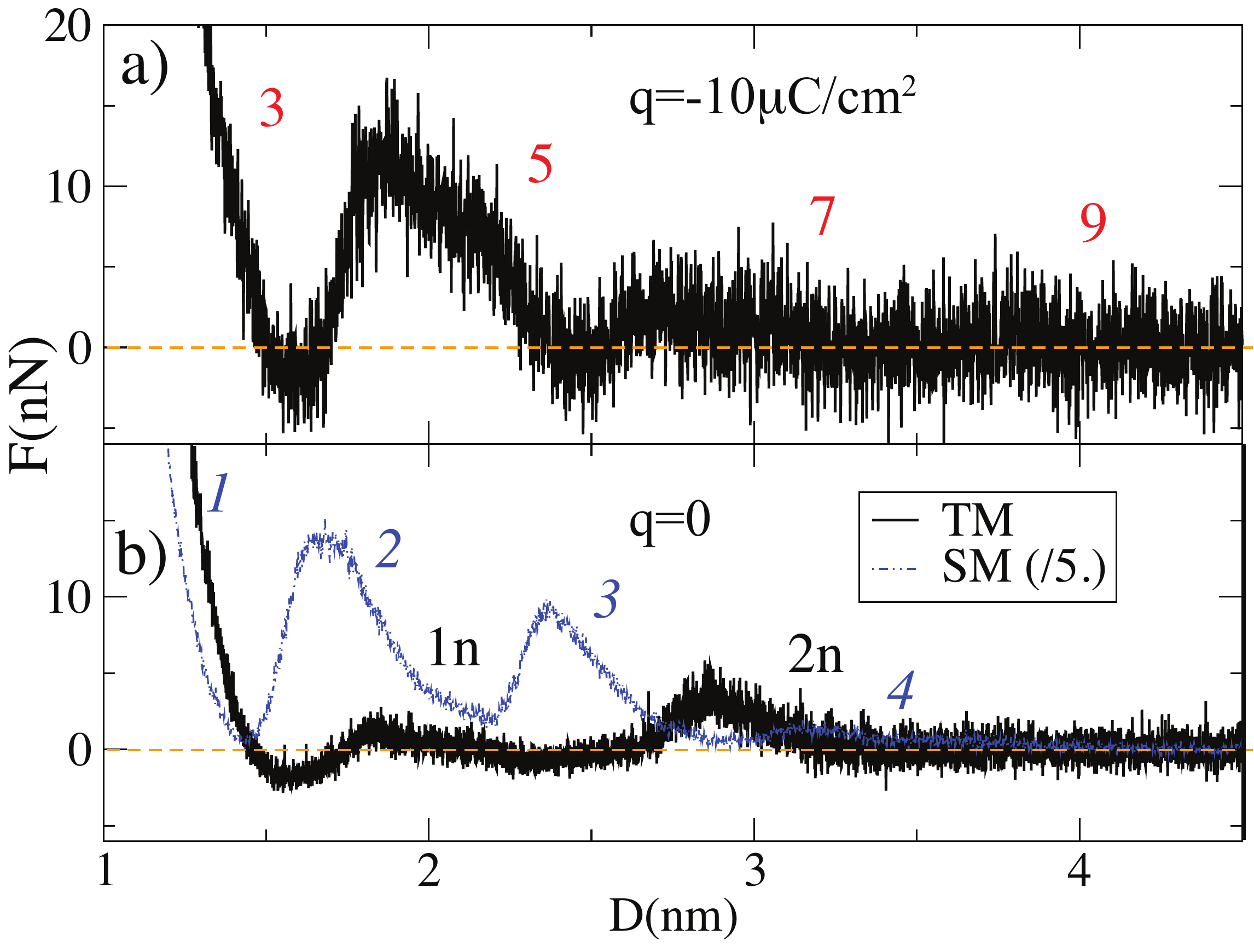}
\caption{(Color online) a) Force-distance curve for the TM liquid and a charge density 
of $q=-10 \mu C/cm^2$ on both plates.
b) Force-distance curves with neutral plates for the TM (black) and SM (blue) liquid.}
\label{force_dist_ch0}
\end{figure}

\begin{figure}
\includegraphics[width=1.\linewidth]{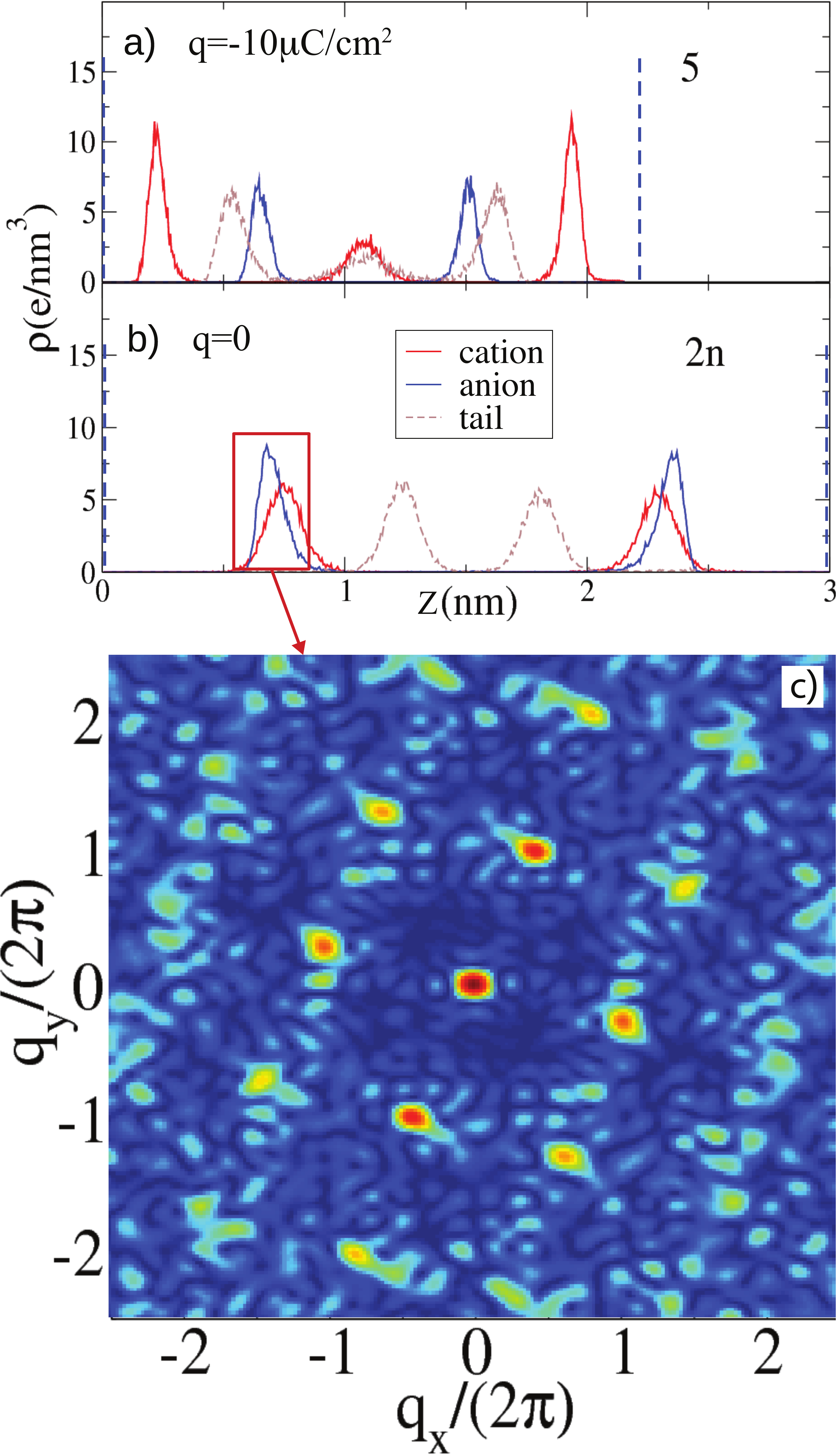}
\caption{(Color online) a) Density profile between plates for the TM liquid
forming a stable five layer film between plates charged with $q=-10 \mu C/cm^2$,
pressed together with a force $F=7nN$. b) Density profile for the same system under the
same force, but now with neutral plates. The interplate gap has collapsed, and much of the IL
has been squeezed out.  Cations and anions have merged into a single neutral layer,
bound to the plates by short-range adhesion. The cation tails point outward and form a weakly repulsive
coating. Vertical dashed lines indicate the position of plates. c)  Structure factor $S_{z_0}(\mathbf{q})$ 
for the layer enclosed in the red rectangle,  showing that ions form a rocksalt (100)  square monolayer.}
\label{square}
\end{figure}

\begin{figure}
\includegraphics[width=1.\linewidth]{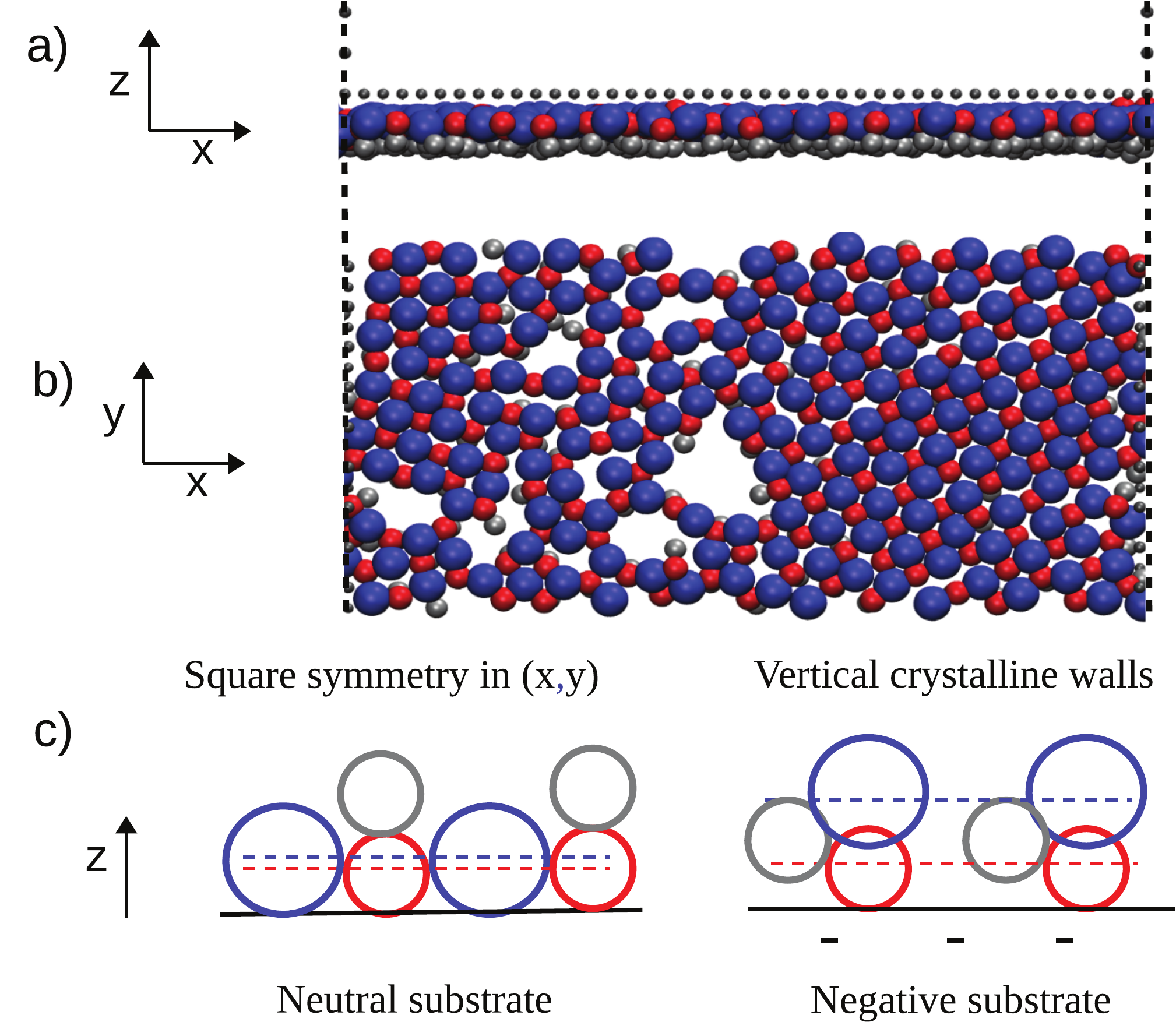}
\caption{(Color online) a) Side view of TM liquid in the state $2n$ with neutral plates. Only one layer is shown for clarity.
b) Top view of the single layer in panel a) showing its square arrangement of oppositely charged ions.
c) Sketch of plate charging induced structural transition from a square  to a triangular planar arrangement.}
\label{pictorial}
\end{figure}

The properties of plate-confined ILs described so far for neutral or nearly neutral plates are strongly 
disturbed once the plates are electrically charged. We study first the 
structural and squeeezout 
behavior under symmetrically charged plates, that is both plates with same charge density $q$. 
In order to preserve the overall system neutrality in simulation, we simultaneously remove from the liquid a number of ions  
amounting to the compensating charge $-2q$. Upon increasing negative $q$ on both plates, the confined IL 
acquired a more solid-like structure, which is mechanically stronger and can bear higher loads before being squeezed out.
Fig.\ref{force_dist_ch0}a shows a force-distance curve for the TM liquid, with a charge density 
of $q=-10 \mu C/cm^2$ on both plates. The force-distance curve show much higher peaks than for neutral plates as in  
in Fig.\ref{force_dist}, signaling the larger force needed for squeezout from $N$ to $N-2$ layers.\\ 
The SM has a very similar behavior, with even higher peaks in the force-distance curve (not shown).  
Fig.\ref{square}a shows the density profile along the z direction for five TM confined layers
between negative plates. Cations crowd up near the negatively 
charged plates, their tail protruding away from the plates. 
For totally neutral plates however, in presence of the sole adhesion forces between 
ions and plates, the structure of the two model liquids differs rather drastically. 
The force-distance curve for the more realistic TM liquid displays just two peaks and no 
evidence for extended layering (see Fig.\ref{force_dist_ch0}b). With neutral plates 
the IL cannot support the 1 nN load down to a gap width of $D\simeq 3nm$. 
A qualitatively similar picture also emerged in recent experimental investigations showing how surface charges affect the molecular
structure and flow properties of ILs \cite{bou-malham10}.
The transverse density profile  for $q=0$ and $D\simeq 3nm$ 
is shown in Fig.\ref{square}b with formation of just two 
neutral layers (which we denote as 2n) where both anions and cations coexist in 
a single monolayer, which is bound to the plate by short-range adhesion. 
The cation tails point outward from the plate giving rise to a weakly repulsive coating, 
which hinders further layering, as suggested by a side view in Fig.\ref{pictorial}a.   
By further increase of vertical load, one of the two neutral monolayers is further squeezed out 
leaving a single one (indicated with 1n) residually confined between the plates.\\ 

The planar structure of these layers is interesting. Fig.\ref{pictorial}b presents a top view of a single layer for the state 2n.
In contrast to the case of charged plates, where the IL crystallized with wall-like z-oriented planes, 
neutral plates induce ions to form neutral layers with square symmetry as shown by Fig.\ref{square}c.\\
Such charging-induced structural transitions may indeed generally arise from a competition between electrostatic and adhesive forces.  
In our case, the sole adhesion to neutral plates binds both cations and anions to the plate, so that they arrange in a rocksalt  (100) 
monolayer with tails pointing outward. A negative plate instead selectively binds  
cations but not anions, giving rise to an alternation of positive and negative layers with tails parallel 
to the plates.  A pictorial representation of these charge induced structural
transitions is given in Fig.\ref{pictorial}c.\\

The SM IL also undergoes a charge induced structural transition, from a triangular to a square arrangement in presence of
neutral plates. Much more ordered than the TM, it displays extended layering, as show by blue dotted line 
in Fig.\ref{force_dist_ch0}b. Essentially, it solidifies  in a rocksalt crystal with the (100) surface in contact with the neutral plate. 
Each peak of the blue curve in Fig.\ref{force_dist_ch0}b, corresponds
to the squeezout of a single neutral (100) layer.

Noteworthy as they may be, these planar structure features seem presently unaccessible to experiments,
and the development of techniques that could allow their study would be very interesting.
In ordinary electrolytes, voltage-controlled phase transitions at electrode-electrolyte 
interfaces have been proven and studied in past years \cite{kornyshev14}.
\section{Antisymmetric, time-dependent plate charging}
\begin{figure}
\includegraphics[width=1.\linewidth]{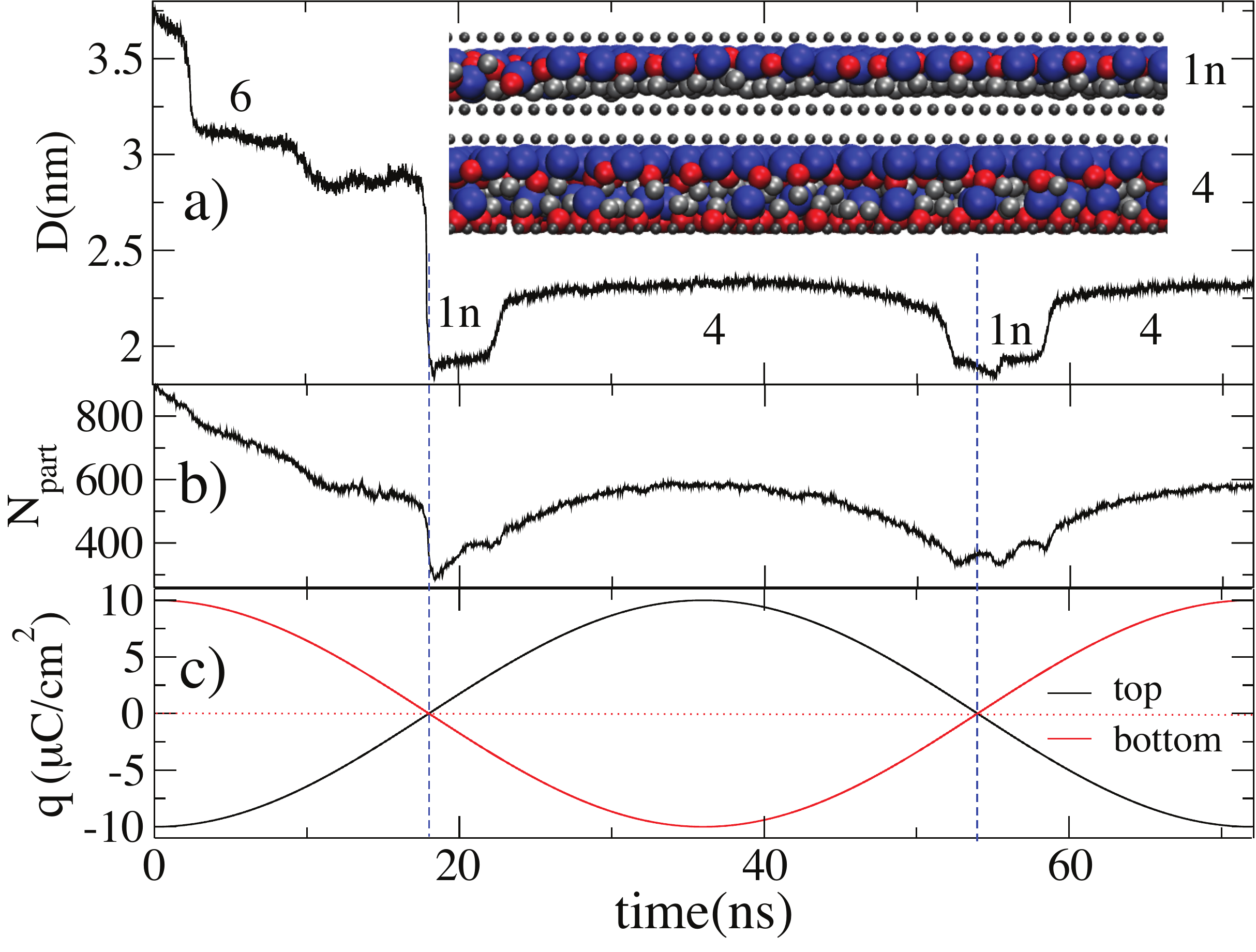}
\caption{(Color online) a) Spontaneous change of interplate distance $D$ between plates as a function of
time for a fixed interplate force $F_n=1nN$ and a slowly oscillating charge with amplitude $Q =10\mu C/cm^2 $ and 
period $\tau=72ns$ (panel c).
b) Number of particles in the interplate gap, $N_{part}$, as a function of time,  under variable charging. 
The time dependent charging causes the plate separation $D$ to open up at maxima (IL suck-in) and to collapse at minima,
(IL squeezout) with a strong electropumping action. The inset in panel a) shows a side view of the confined liquid 
structure for 4 layers ($q=10\mu C/cm^2$)  and for a single neutral layer $1n$ ($q=0$).}
\label{slow}
\end{figure}

In this section we finally study how the order in the TM liquid is affected by opposite charging of the 
two plates. The charging is given a slow sinusoidal variation with time, $q= \pm Q \sin{(2\pi t/\tau)}$ such as would be
caused by externally driven plate charging with a sufficiently low AC frequency to ensure that all the ionic motion
caused inside the interplate gap occurs on a much faster time scale. In this respect, we have checked there is enough 
adiabaticity to ensure that the effects of charge transport and heat dissipation are safely taken care of. Still, as
we shall see, some notable non-adiabatic effects persist, connected with charge-induced solidification.
By keeping the load constant at $F_n=1nN$, 
the charging period $\tau=72ns$ and magnitude $Q= 10\mu C/cm^2 $ we simulate the spontaneous 
evolution of the confined IL and obtain the results shown in Fig.\ref{slow}.

\begin{figure}
\centering
\includegraphics[width=8.5cm,angle=0]{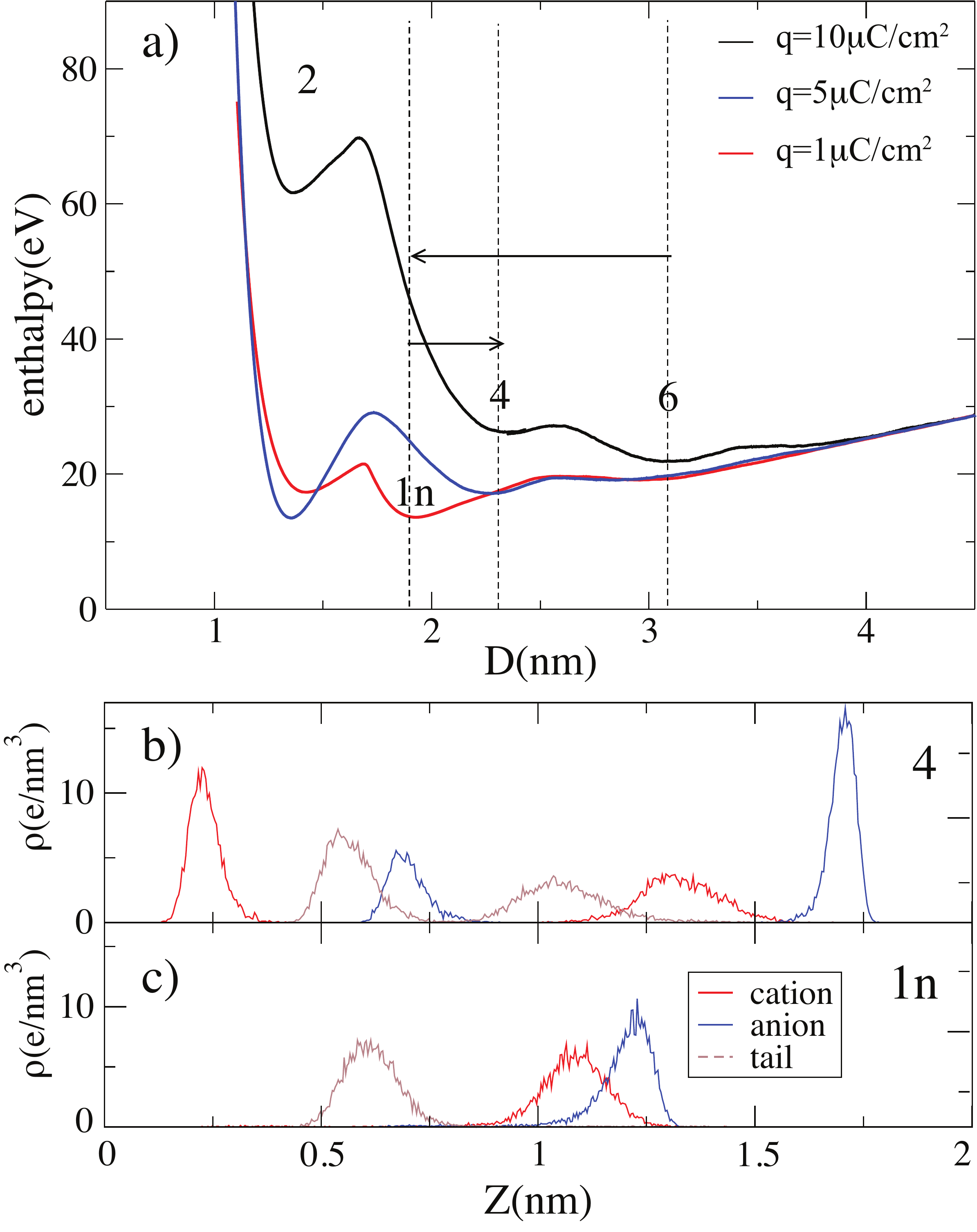}
\caption{(Color online) a) Enthalpy curves for indicated values of charge
density on plates. Numbers indicated in figure denote metastable states at
different number of layers. Peak heights decrease with charge 
and arrows indicate charge induced transitions between the states $6$, $4$ and $1n$.
b) Density profile for the state with $4$ layers. c) Density profile for the state 
with $1$ neutral layer.}
\label{enthalpy}
\end{figure}

Symmetry demands that the equilibrium layering structure for antisymmetric plate charging should occur 
with an even number of players, whereas with symmetric charging the layer number was odd.
The simulation begins with a charge density of $|q(t=0)|= Q= 10\mu C/cm^2$ on plates and the IL structured 
in six layers, which is the stable configuration under the fixed applied load force $F_n=1nN$. 
As the plate charge slowly drops in  
time, the IL structure gradually changes and softens. Eventually, the IL becomes soft enough and squeezout 
suddenly takes place, despite the constant load. The interplate distance $D$ drops, the IL film structure reaching a single, 
charge neutral two-component layer branded $1n$, (see inset of Fig.\ref{slow}a and Fig.\ref{enthalpy}c)  
when $q$=0. That $1n$ structure is a history dependent state that forms asymmetrically depending upon the initial sign of the charge.
As the charging grows again now with inverted sign, the IL is sucked back in and both the gap width $D$ and the number of confined
partcles in the gap, $N_{part}$, grow back to four layers with reversed charge order.
Although four layers is not, as we shall show later, the lowest free enthalpy state, which is instead six layers, 
the metastable four layer state is nearly solid. In that state the kinetics becomes very slow,
so that the time needed to suck back in another bilayer becomes much longer than the simulation time. 
As shown in Fig.\ref{slow}a,b the charging driven inter-plate breathing cycles 
continue periodically, and so does the effective electro-pumping of ions in and out of the gap.  
This kind of  charging dependent phenomena based on spontaneous sucking in of the liquid, solidification
at large charge with buildup of a solid-like resistance to squeezout, followed by sudden squeezout 
associated with melting of the confined layers upon charge reversal is very likely a more general 
feature that could occur in a larger class of ILs beyond our simple model.\\

The charging induced transitions just observed in simulation can actually be predicted by the free enthalpy  
of the system,  $W(D) + F_nD$, where $W(D)$ is calculated the same way as earlier by integrating the interplate force 
from infinity to the gap width $D$. Fig. \ref{enthalpy}a shows the free enthalpy curves so obtained
as a function of $D$ for three values of the plate charge density $\pm q$, starting with six layers 
which is the lowest free enthalpy configuration at a charge density $|q|=10 \mu C/cm^2$.   
Calculated curves show that upon decreasing $|q|$ the energy barriers between states at 
different number of layers decrease gradually, eventually pushing the metastable six-layer state 
leftward to a much smaller distance $D$ between plates. A dramatic squeezout event takes place 
near $q\simeq 0$ (see Fig.\ref{slow}a), with a jump to the single layer state $1n$. 
The successive reversal and increase of plate charge provokes the opposite transition at $q\simeq |3| \mu C/cm^2$
where, as described in the previous section, the IL is sucked in, and structures up into four layers. As the charge 
rises again to $|q|=10 \mu C/cm^2$, the IL never returns to the equilibrium six layer state, indicating that 
the thermal fluctuations and the simulation time are not sufficient to negotiate the four-to-six-layer 
free enthalpy barrier (see Fig.\ref{enthalpy}a black curve).
The nearly solid four layer state effectively resists the insertion of the last bilayer, and blocks the system in a metastable
state which is very long-lived, at least on our simulation time scale. The six-layer equilibrium state should of course be recovered
in a sufficiently slow charge dynamics. \\
 
Similarly to the TM, the SM IL shows even more dramatic  charge induced squeezouts, followed by successive relayerings.
The occurrence of  dramatic electro-squeezout occurring under constant load due to the charge-induced boundary solidification
of the  IL and its sudden melting when the charge is removed, is quite likely a general characteristics that should be explored
in real ILs.
\section{Voltage behavior between plates}

\begin{figure}
\centering
\includegraphics[width=8.5cm,angle=0]{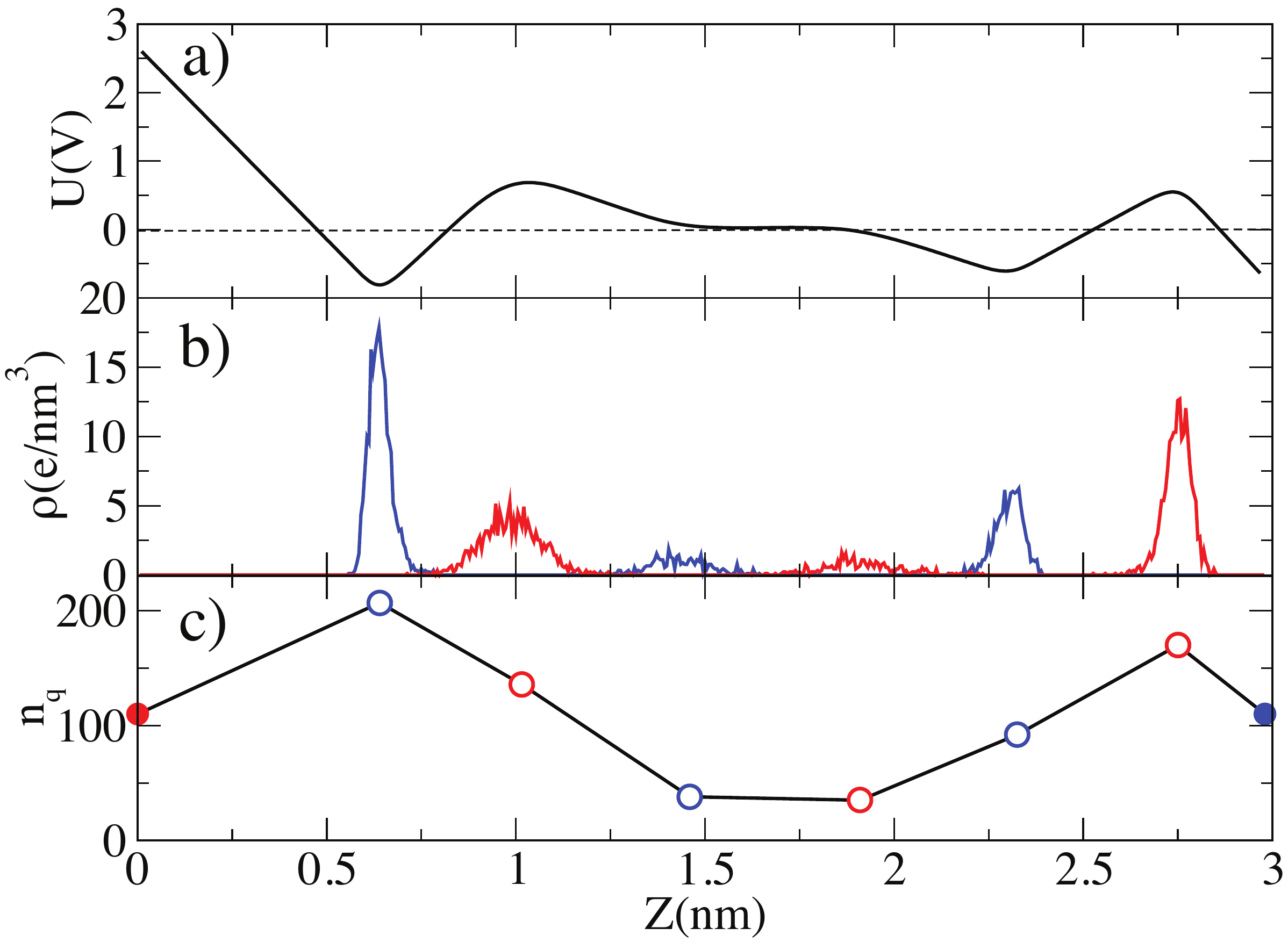}
\caption{(Color online) Voltage (a) and charge profile (b) of the TM, in equilibrium with oppositely charged plates,
$|q|=10 \mu C/cm^2$, force $F_n=1nN$, and a background dielectric constant $\epsilon_r$= 2. The voltage zero is  arbitrarily set at the average value. The voltage drop is about 10 times smaller than that of the two plates in vacuum. 
Note the overscreening near both plates. Panel c) shows the average number 
of charges per layer, $n_l=Q_l/e$, with $Q_l$ the total charge in layer $l$ and  
$e$ the electron charge.}
\label{volgate}
\end{figure}
The plate-confined IL also has interesting capacitive properties. 
The antisymmetric plate charging described in the previous section
corresponds to the application of a voltage between the plates.
Assuming a uniform charge density in the $(x,y)$ plane, and integrating 
Poisson's equation along the vertical axis we obtain an analytical
expression for the voltage as a function of the coordinate $z$ between the plates:
\begin{equation}
U(z)  = -\int_0^z \frac{\rho(z^{'})}{\epsilon_0 \epsilon_r}(z-z^{'}) dz^{'}
+\frac{q}{\epsilon_0\ \epsilon_r} z
\end{equation}
where $\epsilon_0$ is the dielectric permeability of vacuum,  $\rho(z)$ is the density along vertical axis $z$, and the second 
term is the contribution from surface charge density, $\pm |q|$, on the plates. 
Numerical values of $\rho(z)$ calculated from the simulation are integrated 
to obtain the voltage profile, $U(z)$. Fig.\ref{volgate}a
shows the variation of voltage for a charge density $q= \pm 10\mu C/cm^2$
and six IL layers confined, in full thermodynamic equilibrium, between the plates. 
We find that while the potential drop in vacuum would be $\Delta U_V=34 V$, while in 
the presence of TM liquid $\Delta U_{TM}\simeq 3.5V$. Moreover, $\Delta U_{TM}$ remains 
nearly constant near the center of the film, showing that the TM liquid effectively screens 
the plate charge at short distance, as expected from the short Debye screening length of the bulk TM.
Fig.\ref{volgate}b is the density profile along vertical axis, while 
panel \ref{volgate}c displays the average total charge $n_q$ per layer.
The color of circles refers to cations (red) and anions (blue), while 
the full circles indicate the plate charge 
in terms of electron charge $qA/e$, with $A$ area of plate. The
total amount of charge, including the plates and confined liquid, is zero, preserving neutrality. 
The behavior $n_q$ vs $z$ shows a clear overscreening, a phenomenon well known in electrowetting~\cite{kornyshev}
with the charge  of layers in contact with plates about twice larger than that on the plates.   
\section{Discussion and conclusions}

We have undertaken a detailed simulation study of the behavior of simple model ionic liquids confined between 
plates that are being pressed together producing squeezout.  We studied that without and with  electrical charging of the plates, 
showing that charging influences both the structure of the liquid and its squeezout behavior.  Odd-number layering 
and near solidification is found between equally charged plates, 
from where squeezout occurs by successive expulsion of neutral bilayers, as is commonly observed in experiments. The intimate 
structure of the confined film is analysed, showing interesting types of planar order besides the known transverse layering order.  
To investigate the effect of plate charging with opposite sign, where the layering switches from odd to even, we  
carried out slow dynamics simulations where the distance between oppositely charged plates is very slowly and periodically 
varied as a function of time.  Electrically driven squeezout and suck-in transitions are found and shown to give 
rise to a peculiarly fast electro-pumping, which works at constant applied load force, and is simply caused by the charging-induced 
solidification and melting of the IL.   These phenomena are  fully explained thermodynamically in terms of transitions 
between free enthalpy minima,  which as shown  by explicit calculations based on force integration are directly charging-dependent. 

Although the simplicity of the models permits only a limited connection with existing data, the present study 
demonstrates a variety of phenomena that can and will take place when real ionic liquids are confined under 
charged plates. The striking effectiveness of the charging induced solidification-melting process with consequent 
"electroequeezing" discovered in simulation might have practical applications.

Finally, the strong control exerted on the IL nanostructure by the charging of plates suggests that the lubrication 
of plate sliding sliding should be equally affected. That will be the subject of our forthcoming study.   

\section{Acknowledgments}


The authors are grateful to D. Passerone and C. Pignedoli of EMPA (D\"ubendorf, CH) 
for the computational resources and the technical assistance provided, and to A. Kornyshev for helpful discussions. 
Work in Trieste was sponsored by ERC Advanced  Grant 320796 - MODPHYSFRICT, and in part by Sinergia 
Contract CRSII2$_1$36287/1 and by  COST Action MP1303. 

\bibliography{biblio}

\begin{thebibliography}{38}%
\makeatletter
\providecommand \@ifxundefined [1]{%
 \@ifx{#1\undefined}
}%
\providecommand \@ifnum [1]{%
 \ifnum #1\expandafter \@firstoftwo
 \else \expandafter \@secondoftwo
 \fi
}%
\providecommand \@ifx [1]{%
 \ifx #1\expandafter \@firstoftwo
 \else \expandafter \@secondoftwo
 \fi
}%
\providecommand \natexlab [1]{#1}%
\providecommand \enquote  [1]{``#1''}%
\providecommand \bibnamefont  [1]{#1}%
\providecommand \bibfnamefont [1]{#1}%
\providecommand \citenamefont [1]{#1}%
\providecommand \href@noop [0]{\@secondoftwo}%
\providecommand \href [0]{\begingroup \@sanitize@url \@href}%
\providecommand \@href[1]{\@@startlink{#1}\@@href}%
\providecommand \@@href[1]{\endgroup#1\@@endlink}%
\providecommand \@sanitize@url [0]{\catcode `\\12\catcode `\$12\catcode
  `\&12\catcode `\#12\catcode `\^12\catcode `\_12\catcode `\%12\relax}%
\providecommand \@@startlink[1]{}%
\providecommand \@@endlink[0]{}%
\providecommand \url  [0]{\begingroup\@sanitize@url \@url }%
\providecommand \@url [1]{\endgroup\@href {#1}{\urlprefix }}%
\providecommand \urlprefix  [0]{URL }%
\providecommand \Eprint [0]{\href }%
\providecommand \doibase [0]{http://dx.doi.org/}%
\providecommand \selectlanguage [0]{\@gobble}%
\providecommand \bibinfo  [0]{\@secondoftwo}%
\providecommand \bibfield  [0]{\@secondoftwo}%
\providecommand \translation [1]{[#1]}%
\providecommand \BibitemOpen [0]{}%
\providecommand \bibitemStop [0]{}%
\providecommand \bibitemNoStop [0]{.\EOS\space}%
\providecommand \EOS [0]{\spacefactor3000\relax}%
\providecommand \BibitemShut  [1]{\csname bibitem#1\endcsname}%
\let\auto@bib@innerbib\@empty
\bibitem [{\citenamefont {Zhang}\ \emph {et~al.}(2006)\citenamefont {Zhang},
  \citenamefont {Sun}, \citenamefont {He}, \citenamefont {Lu},\ and\
  \citenamefont {Zhang}}]{zhang}%
  \BibitemOpen
  \bibfield  {author} {\bibinfo {author} {\bibfnamefont {S.}~\bibnamefont
  {Zhang}}, \bibinfo {author} {\bibfnamefont {N.}~\bibnamefont {Sun}}, \bibinfo
  {author} {\bibfnamefont {X.}~\bibnamefont {He}}, \bibinfo {author}
  {\bibfnamefont {X.}~\bibnamefont {Lu}}, \ and\ \bibinfo {author}
  {\bibfnamefont {X.}~\bibnamefont {Zhang}},\ }\href@noop {} {\bibfield
  {journal} {\bibinfo  {journal} {J. Phys. Chem. Ref. Data}\ }\textbf {\bibinfo
  {volume} {35}},\ \bibinfo {pages} {1475} (\bibinfo {year}
  {2006})}\BibitemShut {NoStop}%
\bibitem [{\citenamefont {Greaves}\ \emph {et~al.}(2006)\citenamefont
  {Greaves}, \citenamefont {Weerawardena}, \citenamefont {Fong}, \citenamefont
  {Krodkiewska},\ and\ \citenamefont {Drummond}}]{greaves06}%
  \BibitemOpen
  \bibfield  {author} {\bibinfo {author} {\bibfnamefont {T.~L.}\ \bibnamefont
  {Greaves}}, \bibinfo {author} {\bibfnamefont {A.}~\bibnamefont
  {Weerawardena}}, \bibinfo {author} {\bibfnamefont {C.}~\bibnamefont {Fong}},
  \bibinfo {author} {\bibfnamefont {I.}~\bibnamefont {Krodkiewska}}, \ and\
  \bibinfo {author} {\bibfnamefont {C.~J.}\ \bibnamefont {Drummond}},\
  }\href@noop {} {\bibfield  {journal} {\bibinfo  {journal} {J. Phys. Chem. B}\
  }\textbf {\bibinfo {volume} {110}},\ \bibinfo {pages} {22479} (\bibinfo
  {year} {2006})}\BibitemShut {NoStop}%
\bibitem [{\citenamefont {Plechkova}\ and\ \citenamefont
  {Seddon}(2008)}]{plechkova08}%
  \BibitemOpen
  \bibfield  {author} {\bibinfo {author} {\bibfnamefont {N.~V.}\ \bibnamefont
  {Plechkova}}\ and\ \bibinfo {author} {\bibfnamefont {K.~R.}\ \bibnamefont
  {Seddon}},\ }\href@noop {} {\bibfield  {journal} {\bibinfo  {journal} {Chem.
  Soc. Rev.}\ }\textbf {\bibinfo {volume} {37}},\ \bibinfo {pages} {123}
  (\bibinfo {year} {2008})}\BibitemShut {NoStop}%
\bibitem [{\citenamefont {Mezger}\ \emph {et~al.}(2008)\citenamefont {Mezger},
  \citenamefont {Schr\"{o}der}, \citenamefont {Reichert}, \citenamefont
  {Schramm}, \citenamefont {Okasinski}, \citenamefont {Sch\"{o}der},
  \citenamefont {Honkim\"{a}ki}, \citenamefont {Deutsch}, \citenamefont {Ocko},
  \citenamefont {Ralston}, \citenamefont {Rohwerder}, \citenamefont
  {Stratmann},\ and\ \citenamefont {Dosch}}]{mezger08}%
  \BibitemOpen
  \bibfield  {author} {\bibinfo {author} {\bibfnamefont {M.}~\bibnamefont
  {Mezger}}, \bibinfo {author} {\bibfnamefont {H.}~\bibnamefont
  {Schr\"{o}der}}, \bibinfo {author} {\bibfnamefont {H.}~\bibnamefont
  {Reichert}}, \bibinfo {author} {\bibfnamefont {S.}~\bibnamefont {Schramm}},
  \bibinfo {author} {\bibfnamefont {J.~S.}\ \bibnamefont {Okasinski}}, \bibinfo
  {author} {\bibfnamefont {S.}~\bibnamefont {Sch\"{o}der}}, \bibinfo {author}
  {\bibfnamefont {V.}~\bibnamefont {Honkim\"{a}ki}}, \bibinfo {author}
  {\bibfnamefont {M.}~\bibnamefont {Deutsch}}, \bibinfo {author} {\bibfnamefont
  {B.~M.}\ \bibnamefont {Ocko}}, \bibinfo {author} {\bibfnamefont
  {J.}~\bibnamefont {Ralston}}, \bibinfo {author} {\bibfnamefont
  {M.}~\bibnamefont {Rohwerder}}, \bibinfo {author} {\bibfnamefont
  {M.}~\bibnamefont {Stratmann}}, \ and\ \bibinfo {author} {\bibfnamefont
  {H.}~\bibnamefont {Dosch}},\ }\href@noop {} {\bibfield  {journal} {\bibinfo
  {journal} {Science}\ }\textbf {\bibinfo {volume} {322}},\ \bibinfo {pages}
  {424} (\bibinfo {year} {2008})}\BibitemShut {NoStop}%
\bibitem [{\citenamefont {Hayes}\ \emph {et~al.}(2011)\citenamefont {Hayes},
  \citenamefont {Borisenko}, \citenamefont {Tam}, \citenamefont {Howlett},
  \citenamefont {Endres},\ and\ \citenamefont {Atkin}}]{hayes11}%
  \BibitemOpen
  \bibfield  {author} {\bibinfo {author} {\bibfnamefont {R.}~\bibnamefont
  {Hayes}}, \bibinfo {author} {\bibfnamefont {N.}~\bibnamefont {Borisenko}},
  \bibinfo {author} {\bibfnamefont {M.~K.}\ \bibnamefont {Tam}}, \bibinfo
  {author} {\bibfnamefont {P.~C.}\ \bibnamefont {Howlett}}, \bibinfo {author}
  {\bibfnamefont {F.}~\bibnamefont {Endres}}, \ and\ \bibinfo {author}
  {\bibfnamefont {R.}~\bibnamefont {Atkin}},\ }\href@noop {} {\bibfield
  {journal} {\bibinfo  {journal} {J. Phys. Chem. C}\ }\textbf {\bibinfo
  {volume} {115}},\ \bibinfo {pages} {6855} (\bibinfo {year}
  {2011})}\BibitemShut {NoStop}%
\bibitem [{\citenamefont {Bazant}\ \emph {et~al.}(2011)\citenamefont {Bazant},
  \citenamefont {Storey},\ and\ \citenamefont {Kornyshev}}]{kornyshevtheory}%
  \BibitemOpen
  \bibfield  {author} {\bibinfo {author} {\bibfnamefont {M.}~\bibnamefont
  {Bazant}}, \bibinfo {author} {\bibfnamefont {B.}~\bibnamefont {Storey}}, \
  and\ \bibinfo {author} {\bibfnamefont {A.}~\bibnamefont {Kornyshev}},\
  }\href@noop {} {\bibfield  {journal} {\bibinfo  {journal} {Phys. Rev. Lett.}\
  }\textbf {\bibinfo {volume} {106}},\ \bibinfo {pages} {046102} (\bibinfo
  {year} {2011})}\BibitemShut {NoStop}%
\bibitem [{\citenamefont {Kornyshev}(2007)}]{kornyshev07}%
  \BibitemOpen
  \bibfield  {author} {\bibinfo {author} {\bibfnamefont {A.~A.}\ \bibnamefont
  {Kornyshev}},\ }\href@noop {} {\bibfield  {journal} {\bibinfo  {journal} {J.
  Phys. Chem. B}\ }\textbf {\bibinfo {volume} {111}},\ \bibinfo {pages} {5545}
  (\bibinfo {year} {2007})}\BibitemShut {NoStop}%
\bibitem [{\citenamefont {Atkin}\ and\ \citenamefont {Warr}(2007)}]{atkin07}%
  \BibitemOpen
  \bibfield  {author} {\bibinfo {author} {\bibfnamefont {R.}~\bibnamefont
  {Atkin}}\ and\ \bibinfo {author} {\bibfnamefont {G.~G.}\ \bibnamefont
  {Warr}},\ }\href@noop {} {\bibfield  {journal} {\bibinfo  {journal} {J. Phys.
  Chem. C}\ }\textbf {\bibinfo {volume} {111}},\ \bibinfo {pages} {5162}
  (\bibinfo {year} {2007})}\BibitemShut {NoStop}%
\bibitem [{\citenamefont {Sweeney}\ \emph {et~al.}(2012)\citenamefont
  {Sweeney}, \citenamefont {Hausen}, \citenamefont {Hayes}, \citenamefont
  {Webber}, \citenamefont {Endres}, \citenamefont {Rutland}, \citenamefont
  {Bennewitz},\ and\ \citenamefont {Atkin}}]{bennewitz}%
  \BibitemOpen
  \bibfield  {author} {\bibinfo {author} {\bibfnamefont {J.}~\bibnamefont
  {Sweeney}}, \bibinfo {author} {\bibfnamefont {F.}~\bibnamefont {Hausen}},
  \bibinfo {author} {\bibfnamefont {R.}~\bibnamefont {Hayes}}, \bibinfo
  {author} {\bibfnamefont {G.~B.}\ \bibnamefont {Webber}}, \bibinfo {author}
  {\bibfnamefont {F.}~\bibnamefont {Endres}}, \bibinfo {author} {\bibfnamefont
  {M.~W.}\ \bibnamefont {Rutland}}, \bibinfo {author} {\bibfnamefont
  {R.}~\bibnamefont {Bennewitz}}, \ and\ \bibinfo {author} {\bibfnamefont
  {R.}~\bibnamefont {Atkin}},\ }\href@noop {} {\bibfield  {journal} {\bibinfo
  {journal} {Phys. Rev. Lett.}\ }\textbf {\bibinfo {volume} {109}},\ \bibinfo
  {pages} {155502} (\bibinfo {year} {2012})}\BibitemShut {NoStop}%
\bibitem [{\citenamefont {Smith}\ \emph
  {et~al.}(2013{\natexlab{a}})\citenamefont {Smith}, \citenamefont {Lovelock},
  \citenamefont {Gosvami}, \citenamefont {Welton},\ and\ \citenamefont
  {Perkin}}]{perkin}%
  \BibitemOpen
  \bibfield  {author} {\bibinfo {author} {\bibfnamefont {A.}~\bibnamefont
  {Smith}}, \bibinfo {author} {\bibfnamefont {K.}~\bibnamefont {Lovelock}},
  \bibinfo {author} {\bibfnamefont {N.}~\bibnamefont {Gosvami}}, \bibinfo
  {author} {\bibfnamefont {T.}~\bibnamefont {Welton}}, \ and\ \bibinfo {author}
  {\bibfnamefont {S.}~\bibnamefont {Perkin}},\ }\href@noop {} {\bibfield
  {journal} {\bibinfo  {journal} {Phys. Chem. Chem. Phys.}\ }\textbf {\bibinfo
  {volume} {15}},\ \bibinfo {pages} {15317} (\bibinfo {year}
  {2013}{\natexlab{a}})}\BibitemShut {NoStop}%
\bibitem [{\citenamefont {Li}\ \emph {et~al.}(2014)\citenamefont {Li},
  \citenamefont {Wood}, \citenamefont {Rutland},\ and\ \citenamefont
  {Atkin}}]{li14}%
  \BibitemOpen
  \bibfield  {author} {\bibinfo {author} {\bibfnamefont {H.}~\bibnamefont
  {Li}}, \bibinfo {author} {\bibfnamefont {R.~J.}\ \bibnamefont {Wood}},
  \bibinfo {author} {\bibfnamefont {M.~W.}\ \bibnamefont {Rutland}}, \ and\
  \bibinfo {author} {\bibfnamefont {R.}~\bibnamefont {Atkin}},\ }\href@noop {}
  {\bibfield  {journal} {\bibinfo  {journal} {Chem. Commun.}\ }\textbf
  {\bibinfo {volume} {50}},\ \bibinfo {pages} {4368} (\bibinfo {year}
  {2014})}\BibitemShut {NoStop}%
\bibitem [{\citenamefont {Mendonc\c{a}}\ \emph {et~al.}(2013)\citenamefont
  {Mendonc\c{a}}, \citenamefont {P\`{a}dua},\ and\ \citenamefont
  {Malfreyt}}]{padua13}%
  \BibitemOpen
  \bibfield  {author} {\bibinfo {author} {\bibfnamefont {A.}~\bibnamefont
  {Mendonc\c{a}}}, \bibinfo {author} {\bibfnamefont {A.}~\bibnamefont
  {P\`{a}dua}}, \ and\ \bibinfo {author} {\bibfnamefont {P.}~\bibnamefont
  {Malfreyt}},\ }\href@noop {} {\bibfield  {journal} {\bibinfo  {journal} {J.
  Chem. Theory Comput.}\ }\textbf {\bibinfo {volume} {9}},\ \bibinfo {pages}
  {1600} (\bibinfo {year} {2013})}\BibitemShut {NoStop}%
\bibitem [{\citenamefont {Canova}\ \emph {et~al.}(2014)\citenamefont {Canova},
  \citenamefont {Matsubara}, \citenamefont {Mizukami}, \citenamefont
  {Kurihara},\ and\ \citenamefont {Shluger}}]{federici14}%
  \BibitemOpen
  \bibfield  {author} {\bibinfo {author} {\bibfnamefont {F.~F.}\ \bibnamefont
  {Canova}}, \bibinfo {author} {\bibfnamefont {H.}~\bibnamefont {Matsubara}},
  \bibinfo {author} {\bibfnamefont {M.}~\bibnamefont {Mizukami}}, \bibinfo
  {author} {\bibfnamefont {K.}~\bibnamefont {Kurihara}}, \ and\ \bibinfo
  {author} {\bibfnamefont {A.~L.}\ \bibnamefont {Shluger}},\ }\href@noop {}
  {\bibfield  {journal} {\bibinfo  {journal} {Phys. Chem. Chem. Phys.}\
  }\textbf {\bibinfo {volume} {16}},\ \bibinfo {pages} {8247} (\bibinfo {year}
  {2014})}\BibitemShut {NoStop}%
\bibitem [{\citenamefont {Fedorov}\ and\ \citenamefont
  {Kornyshev}(2008)}]{kornyshev}%
  \BibitemOpen
  \bibfield  {author} {\bibinfo {author} {\bibfnamefont {M.}~\bibnamefont
  {Fedorov}}\ and\ \bibinfo {author} {\bibfnamefont {A.}~\bibnamefont
  {Kornyshev}},\ }\href@noop {} {\bibfield  {journal} {\bibinfo  {journal} {J.
  Phys. Chem. B}\ }\textbf {\bibinfo {volume} {112}},\ \bibinfo {pages} {11868}
  (\bibinfo {year} {2008})}\BibitemShut {NoStop}%
\bibitem [{\citenamefont {Wang}\ and\ \citenamefont {Priest}(2013)}]{wang}%
  \BibitemOpen
  \bibfield  {author} {\bibinfo {author} {\bibfnamefont {Z.}~\bibnamefont
  {Wang}}\ and\ \bibinfo {author} {\bibfnamefont {C.}~\bibnamefont {Priest}},\
  }\href@noop {} {\bibfield  {journal} {\bibinfo  {journal} {Langmuir}\
  }\textbf {\bibinfo {volume} {29}},\ \bibinfo {pages} {11344} (\bibinfo {year}
  {2013})}\BibitemShut {NoStop}%
\bibitem [{\citenamefont {Fumi}\ and\ \citenamefont {Tosi}(1964)}]{fumi}%
  \BibitemOpen
  \bibfield  {author} {\bibinfo {author} {\bibfnamefont {F.~G.}\ \bibnamefont
  {Fumi}}\ and\ \bibinfo {author} {\bibfnamefont {M.~P.}\ \bibnamefont
  {Tosi}},\ }\href@noop {} {\bibfield  {journal} {\bibinfo  {journal} {J. Phys.
  Chem. Solids}\ }\textbf {\bibinfo {volume} {25}},\ \bibinfo {pages} {31}
  (\bibinfo {year} {1964})}\BibitemShut {NoStop}%
\bibitem [{\citenamefont {Gonz\'{a}lez-Melchor}\ \emph
  {et~al.}(2005)\citenamefont {Gonz\'{a}lez-Melchor}, \citenamefont {Bresme},\
  and\ \citenamefont {Alejandre}}]{alejandre}%
  \BibitemOpen
  \bibfield  {author} {\bibinfo {author} {\bibfnamefont {M.}~\bibnamefont
  {Gonz\'{a}lez-Melchor}}, \bibinfo {author} {\bibfnamefont {F.}~\bibnamefont
  {Bresme}}, \ and\ \bibinfo {author} {\bibfnamefont {J.}~\bibnamefont
  {Alejandre}},\ }\href@noop {} {\bibfield  {journal} {\bibinfo  {journal} {J.
  Chem. Phys.}\ }\textbf {\bibinfo {volume} {122}},\ \bibinfo {pages} {104710}
  (\bibinfo {year} {2005})}\BibitemShut {NoStop}%
\bibitem [{\citenamefont {Plimpton}(1995)}]{lammps}%
  \BibitemOpen
  \bibfield  {author} {\bibinfo {author} {\bibfnamefont {S.}~\bibnamefont
  {Plimpton}},\ }\href@noop {} {\bibfield  {journal} {\bibinfo  {journal} {J.
  Comp. Phys.}\ }\textbf {\bibinfo {volume} {117}},\ \bibinfo {pages} {1}
  (\bibinfo {year} {1995})}\BibitemShut {NoStop}%
\bibitem [{\citenamefont {Beattie}\ \emph {et~al.}(2013)\citenamefont
  {Beattie}, \citenamefont {Espinosa-Marzal}, \citenamefont {Tracey T. M.~Ho},
  \citenamefont {Ralston}, \citenamefont {Richard}, \citenamefont
  {Sellapperumage},\ and\ \citenamefont {Krasowska}}]{beattie2013}%
  \BibitemOpen
  \bibfield  {author} {\bibinfo {author} {\bibfnamefont {D.~A.}\ \bibnamefont
  {Beattie}}, \bibinfo {author} {\bibfnamefont {R.~M.}\ \bibnamefont
  {Espinosa-Marzal}}, \bibinfo {author} {\bibfnamefont {M.~N.~P.}\ \bibnamefont
  {Tracey T. M.~Ho}}, \bibinfo {author} {\bibfnamefont {J.}~\bibnamefont
  {Ralston}}, \bibinfo {author} {\bibfnamefont {C.~J.~E.}\ \bibnamefont
  {Richard}}, \bibinfo {author} {\bibfnamefont {P.~M.~F.}\ \bibnamefont
  {Sellapperumage}}, \ and\ \bibinfo {author} {\bibfnamefont {M.}~\bibnamefont
  {Krasowska}},\ }\href@noop {} {\bibfield  {journal} {\bibinfo  {journal} {J.
  Phys. Chem. C}\ }\textbf {\bibinfo {volume} {117}},\ \bibinfo {pages} {23676}
  (\bibinfo {year} {2013})}\BibitemShut {NoStop}%
\bibitem [{\citenamefont {Welters}\ and\ \citenamefont
  {Fokkink}(1998)}]{welters98}%
  \BibitemOpen
  \bibfield  {author} {\bibinfo {author} {\bibfnamefont {W.~J.~J.}\
  \bibnamefont {Welters}}\ and\ \bibinfo {author} {\bibfnamefont {L.~G.~J.}\
  \bibnamefont {Fokkink}},\ }\href@noop {} {\bibfield  {journal} {\bibinfo
  {journal} {Langmuir}\ }\textbf {\bibinfo {volume} {14}},\ \bibinfo {pages}
  {1535} (\bibinfo {year} {1998})}\BibitemShut {NoStop}%
\bibitem [{\citenamefont {Paneru}\ \emph {et~al.}(2010)\citenamefont {Paneru},
  \citenamefont {Priest}, \citenamefont {Sedev},\ and\ \citenamefont
  {Ralston}}]{paneru10}%
  \BibitemOpen
  \bibfield  {author} {\bibinfo {author} {\bibfnamefont {M.}~\bibnamefont
  {Paneru}}, \bibinfo {author} {\bibfnamefont {C.}~\bibnamefont {Priest}},
  \bibinfo {author} {\bibfnamefont {R.}~\bibnamefont {Sedev}}, \ and\ \bibinfo
  {author} {\bibfnamefont {J.}~\bibnamefont {Ralston}},\ }\href@noop {}
  {\bibfield  {journal} {\bibinfo  {journal} {J. Am. Chem. Soc.}\ }\textbf
  {\bibinfo {volume} {132}},\ \bibinfo {pages} {8301} (\bibinfo {year}
  {2010})}\BibitemShut {NoStop}%
\bibitem [{\citenamefont {Israelachvili}(2011)}]{israelachvili11}%
  \BibitemOpen
  \bibfield  {author} {\bibinfo {author} {\bibfnamefont {J.~N.}\ \bibnamefont
  {Israelachvili}},\ }\href@noop {} {\emph {\bibinfo {title} {Intermolecular
  and surface forces}}},\ \bibinfo {edition} {3rd}\ ed.\ (\bibinfo  {publisher}
  {Academic Press},\ \bibinfo {address} {San Diego, CA},\ \bibinfo {year}
  {2011})\BibitemShut {NoStop}%
\bibitem [{\citenamefont {Persson}\ and\ \citenamefont
  {Tosatti}(1994)}]{persson94}%
  \BibitemOpen
  \bibfield  {author} {\bibinfo {author} {\bibfnamefont {B.~N.~J.}\
  \bibnamefont {Persson}}\ and\ \bibinfo {author} {\bibfnamefont
  {E.}~\bibnamefont {Tosatti}},\ }\href@noop {} {\bibfield  {journal} {\bibinfo
   {journal} {Phys. Rev. B}\ }\textbf {\bibinfo {volume} {50}},\ \bibinfo
  {pages} {5590} (\bibinfo {year} {1994})}\BibitemShut {NoStop}%
\bibitem [{\citenamefont {Gao}\ \emph {et~al.}(1997)\citenamefont {Gao},
  \citenamefont {Luedtke},\ and\ \citenamefont {Landman}}]{gao97b}%
  \BibitemOpen
  \bibfield  {author} {\bibinfo {author} {\bibfnamefont {J.~P.}\ \bibnamefont
  {Gao}}, \bibinfo {author} {\bibfnamefont {W.~D.}\ \bibnamefont {Luedtke}}, \
  and\ \bibinfo {author} {\bibfnamefont {U.}~\bibnamefont {Landman}},\
  }\href@noop {} {\bibfield  {journal} {\bibinfo  {journal} {J. Chem. Phys.}\
  }\textbf {\bibinfo {volume} {106}},\ \bibinfo {pages} {4309} (\bibinfo {year}
  {1997})}\BibitemShut {NoStop}%
\bibitem [{\citenamefont {Tartaglino}\ \emph {et~al.}(2002)\citenamefont
  {Tartaglino}, \citenamefont {Persson}, \citenamefont {Volokitin},\ and\
  \citenamefont {Tosatti}}]{tartaglino02}%
  \BibitemOpen
  \bibfield  {author} {\bibinfo {author} {\bibfnamefont {U.}~\bibnamefont
  {Tartaglino}}, \bibinfo {author} {\bibfnamefont {B.~N.~J.}\ \bibnamefont
  {Persson}}, \bibinfo {author} {\bibfnamefont {A.~I.}\ \bibnamefont
  {Volokitin}}, \ and\ \bibinfo {author} {\bibfnamefont {E.}~\bibnamefont
  {Tosatti}},\ }\href@noop {} {\bibfield  {journal} {\bibinfo  {journal} {Phys.
  Rev. B}\ }\textbf {\bibinfo {volume} {66}},\ \bibinfo {pages} {214207}
  (\bibinfo {year} {2002})}\BibitemShut {NoStop}%
\bibitem [{\citenamefont {Tartaglino}\ \emph {et~al.}(2006)\citenamefont
  {Tartaglino}, \citenamefont {Sivebaek}, \citenamefont {Persson},\ and\
  \citenamefont {Tosatti}}]{tartaglino06}%
  \BibitemOpen
  \bibfield  {author} {\bibinfo {author} {\bibfnamefont {U.}~\bibnamefont
  {Tartaglino}}, \bibinfo {author} {\bibfnamefont {I.~M.}\ \bibnamefont
  {Sivebaek}}, \bibinfo {author} {\bibfnamefont {B.~N.~J.}\ \bibnamefont
  {Persson}}, \ and\ \bibinfo {author} {\bibfnamefont {E.}~\bibnamefont
  {Tosatti}},\ }\href@noop {} {\bibfield  {journal} {\bibinfo  {journal} {J.
  Chem. Phys.}\ }\textbf {\bibinfo {volume} {125}},\ \bibinfo {pages} {014704}
  (\bibinfo {year} {2006})}\BibitemShut {NoStop}%
\bibitem [{\citenamefont {Mugele}\ and\ \citenamefont
  {Salmeron}(2000)}]{mugele00}%
  \BibitemOpen
  \bibfield  {author} {\bibinfo {author} {\bibfnamefont {F.}~\bibnamefont
  {Mugele}}\ and\ \bibinfo {author} {\bibfnamefont {M.}~\bibnamefont
  {Salmeron}},\ }\href@noop {} {\bibfield  {journal} {\bibinfo  {journal}
  {Phys. Rev. Lett.}\ }\textbf {\bibinfo {volume} {84}},\ \bibinfo {pages}
  {5796} (\bibinfo {year} {2000})}\BibitemShut {NoStop}%
\bibitem [{\citenamefont {Zilberman}\ \emph {et~al.}(2001)\citenamefont
  {Zilberman}, \citenamefont {Persson}, \citenamefont {Nitzan}, \citenamefont
  {Mugele},\ and\ \citenamefont {Salmeron}}]{zilberman01}%
  \BibitemOpen
  \bibfield  {author} {\bibinfo {author} {\bibfnamefont {S.}~\bibnamefont
  {Zilberman}}, \bibinfo {author} {\bibfnamefont {B.~N.~J.}\ \bibnamefont
  {Persson}}, \bibinfo {author} {\bibfnamefont {A.}~\bibnamefont {Nitzan}},
  \bibinfo {author} {\bibfnamefont {F.}~\bibnamefont {Mugele}}, \ and\ \bibinfo
  {author} {\bibfnamefont {M.}~\bibnamefont {Salmeron}},\ }\href@noop {}
  {\bibfield  {journal} {\bibinfo  {journal} {Phys. Rev. E}\ }\textbf {\bibinfo
  {volume} {63}},\ \bibinfo {pages} {055103(R)} (\bibinfo {year}
  {2001})}\BibitemShut {NoStop}%
\bibitem [{\citenamefont {Ueno}\ \emph {et~al.}(2010)\citenamefont {Ueno},
  \citenamefont {Kasuya}, \citenamefont {Watanabe}, \citenamefont {Mizukami},\
  and\ \citenamefont {Kurihara}}]{ueno10}%
  \BibitemOpen
  \bibfield  {author} {\bibinfo {author} {\bibfnamefont {K.}~\bibnamefont
  {Ueno}}, \bibinfo {author} {\bibfnamefont {M.}~\bibnamefont {Kasuya}},
  \bibinfo {author} {\bibfnamefont {M.}~\bibnamefont {Watanabe}}, \bibinfo
  {author} {\bibfnamefont {M.}~\bibnamefont {Mizukami}}, \ and\ \bibinfo
  {author} {\bibfnamefont {K.}~\bibnamefont {Kurihara}},\ }\href@noop {}
  {\bibfield  {journal} {\bibinfo  {journal} {Phys. Chem. Chem. Phys.}\
  }\textbf {\bibinfo {volume} {12}},\ \bibinfo {pages} {4066} (\bibinfo {year}
  {2010})}\BibitemShut {NoStop}%
\bibitem [{\citenamefont {Smith}\ \emph
  {et~al.}(2013{\natexlab{b}})\citenamefont {Smith}, \citenamefont {Lovelock},
  \citenamefont {Gosvami}, \citenamefont {Licence}, \citenamefont {Dolan},
  \citenamefont {Welton},\ and\ \citenamefont {Perkin}}]{smith13}%
  \BibitemOpen
  \bibfield  {author} {\bibinfo {author} {\bibfnamefont {A.~M.}\ \bibnamefont
  {Smith}}, \bibinfo {author} {\bibfnamefont {K.~R.~J.}\ \bibnamefont
  {Lovelock}}, \bibinfo {author} {\bibfnamefont {N.~N.}\ \bibnamefont
  {Gosvami}}, \bibinfo {author} {\bibfnamefont {P.}~\bibnamefont {Licence}},
  \bibinfo {author} {\bibfnamefont {A.}~\bibnamefont {Dolan}}, \bibinfo
  {author} {\bibfnamefont {T.}~\bibnamefont {Welton}}, \ and\ \bibinfo {author}
  {\bibfnamefont {S.}~\bibnamefont {Perkin}},\ }\href@noop {} {\bibfield
  {journal} {\bibinfo  {journal} {J. Phys. Chem. Lett.}\ }\textbf {\bibinfo
  {volume} {4}},\ \bibinfo {pages} {378} (\bibinfo {year}
  {2013}{\natexlab{b}})}\BibitemShut {NoStop}%
\bibitem [{\citenamefont {Atkin}\ \emph {et~al.}(2009)\citenamefont {Atkin},
  \citenamefont {Abedin}, \citenamefont {Hayes}, \citenamefont {Gasparotto},
  \citenamefont {Borisenko},\ and\ \citenamefont {Endres}}]{atkin09}%
  \BibitemOpen
  \bibfield  {author} {\bibinfo {author} {\bibfnamefont {R.}~\bibnamefont
  {Atkin}}, \bibinfo {author} {\bibfnamefont {S.~Z.~E.}\ \bibnamefont
  {Abedin}}, \bibinfo {author} {\bibfnamefont {R.}~\bibnamefont {Hayes}},
  \bibinfo {author} {\bibfnamefont {L.~H.~S.}\ \bibnamefont {Gasparotto}},
  \bibinfo {author} {\bibfnamefont {N.}~\bibnamefont {Borisenko}}, \ and\
  \bibinfo {author} {\bibfnamefont {F.}~\bibnamefont {Endres}},\ }\href@noop {}
  {\bibfield  {journal} {\bibinfo  {journal} {J. Phys. Chem. C}\ }\textbf
  {\bibinfo {volume} {113}},\ \bibinfo {pages} {13266} (\bibinfo {year}
  {2009})}\BibitemShut {NoStop}%
\bibitem [{\citenamefont {Hoth}\ \emph {et~al.}(2014)\citenamefont {Hoth},
  \citenamefont {Hausen}, \citenamefont {M\"{u}ser},\ and\ \citenamefont
  {Bennewitz}}]{hoth14}%
  \BibitemOpen
  \bibfield  {author} {\bibinfo {author} {\bibfnamefont {J.}~\bibnamefont
  {Hoth}}, \bibinfo {author} {\bibfnamefont {F.}~\bibnamefont {Hausen}},
  \bibinfo {author} {\bibfnamefont {M.~H.}\ \bibnamefont {M\"{u}ser}}, \ and\
  \bibinfo {author} {\bibfnamefont {R.}~\bibnamefont {Bennewitz}},\ }\href@noop
  {} {\bibfield  {journal} {\bibinfo  {journal} {J. Phys.: Condens. Matter}\
  }\textbf {\bibinfo {volume} {26}},\ \bibinfo {pages} {284110} (\bibinfo
  {year} {2014})}\BibitemShut {NoStop}%
\bibitem [{\citenamefont {Yongsheng~Leng}\ and\ \citenamefont
  {Rao}(2013)}]{leng13}%
  \BibitemOpen
  \bibfield  {author} {\bibinfo {author} {\bibfnamefont {Y.~L.}\ \bibnamefont
  {Yongsheng~Leng}, \bibfnamefont {Yuan~Xiang}}\ and\ \bibinfo {author}
  {\bibfnamefont {Q.}~\bibnamefont {Rao}},\ }\href@noop {} {\bibfield
  {journal} {\bibinfo  {journal} {J. Chem. Phys.}\ }\textbf {\bibinfo {volume}
  {139}},\ \bibinfo {pages} {074704} (\bibinfo {year} {2013})}\BibitemShut
  {NoStop}%
\bibitem [{\citenamefont {Borisenko}\ \emph {et~al.}(2006)\citenamefont
  {Borisenko}, \citenamefont {Abedin},\ and\ \citenamefont
  {Endres}}]{borisenko06}%
  \BibitemOpen
  \bibfield  {author} {\bibinfo {author} {\bibfnamefont {N.}~\bibnamefont
  {Borisenko}}, \bibinfo {author} {\bibfnamefont {S.~Z.~E.}\ \bibnamefont
  {Abedin}}, \ and\ \bibinfo {author} {\bibfnamefont {F.}~\bibnamefont
  {Endres}},\ }\href@noop {} {\bibfield  {journal} {\bibinfo  {journal} {J.
  Phys. Chem. B}\ }\textbf {\bibinfo {volume} {110}},\ \bibinfo {pages} {6250}
  (\bibinfo {year} {2006})}\BibitemShut {NoStop}%
\bibitem [{\citenamefont {Lin}\ \emph {et~al.}(2003)\citenamefont {Lin},
  \citenamefont {Y.~Wang}, \citenamefont {Yuan}, \citenamefont {Xiang},\ and\
  \citenamefont {Mao}}]{lin03}%
  \BibitemOpen
  \bibfield  {author} {\bibinfo {author} {\bibfnamefont {L.}~\bibnamefont
  {Lin}}, \bibinfo {author} {\bibfnamefont {J.~Y.}\ \bibnamefont {Y.~Wang}},
  \bibinfo {author} {\bibfnamefont {Y.}~\bibnamefont {Yuan}}, \bibinfo {author}
  {\bibfnamefont {J.}~\bibnamefont {Xiang}}, \ and\ \bibinfo {author}
  {\bibfnamefont {B.}~\bibnamefont {Mao}},\ }\href@noop {} {\bibfield
  {journal} {\bibinfo  {journal} {Electrochemistry Communications}\ }\textbf
  {\bibinfo {volume} {5}},\ \bibinfo {pages} {995} (\bibinfo {year}
  {2003})}\BibitemShut {NoStop}%
\bibitem [{\citenamefont {Segura}\ \emph {et~al.}(2013)\citenamefont {Segura},
  \citenamefont {Elbourne}, \citenamefont {Wanless}, \citenamefont {Warr},
  \citenamefont {Vo\"{i}chovsky},\ and\ \citenamefont {Atkin}}]{segura13}%
  \BibitemOpen
  \bibfield  {author} {\bibinfo {author} {\bibfnamefont {J.~J.}\ \bibnamefont
  {Segura}}, \bibinfo {author} {\bibfnamefont {A.}~\bibnamefont {Elbourne}},
  \bibinfo {author} {\bibfnamefont {E.~J.}\ \bibnamefont {Wanless}}, \bibinfo
  {author} {\bibfnamefont {G.~G.}\ \bibnamefont {Warr}}, \bibinfo {author}
  {\bibfnamefont {K.}~\bibnamefont {Vo\"{i}chovsky}}, \ and\ \bibinfo {author}
  {\bibfnamefont {R.}~\bibnamefont {Atkin}},\ }\href@noop {} {\bibfield
  {journal} {\bibinfo  {journal} {Phys. Chem. Chem. Phys.}\ }\textbf {\bibinfo
  {volume} {15}},\ \bibinfo {pages} {3320} (\bibinfo {year}
  {2013})}\BibitemShut {NoStop}%
\bibitem [{\citenamefont {Bou-Malham}\ and\ \citenamefont
  {Bureau}(2010)}]{bou-malham10}%
  \BibitemOpen
  \bibfield  {author} {\bibinfo {author} {\bibfnamefont {I.}~\bibnamefont
  {Bou-Malham}}\ and\ \bibinfo {author} {\bibfnamefont {L.}~\bibnamefont
  {Bureau}},\ }\href@noop {} {\bibfield  {journal} {\bibinfo  {journal} {Soft
  Matter}\ }\textbf {\bibinfo {volume} {6}},\ \bibinfo {pages} {4062} (\bibinfo
  {year} {2010})}\BibitemShut {NoStop}%
\bibitem [{\citenamefont {Kornyshev}\ and\ \citenamefont
  {Qiao}(2014)}]{kornyshev14}%
  \BibitemOpen
  \bibfield  {author} {\bibinfo {author} {\bibfnamefont {A.~A.}\ \bibnamefont
  {Kornyshev}}\ and\ \bibinfo {author} {\bibfnamefont {R.}~\bibnamefont
  {Qiao}},\ }\href@noop {} {\bibfield  {journal} {\bibinfo  {journal} {J. Phys.
  Chem. C}\ }\textbf {\bibinfo {volume} {118}},\ \bibinfo {pages} {18285}
  (\bibinfo {year} {2014})}\BibitemShut {NoStop}%
\end{thebibliography}%

\end{document}